\documentclass[12pt]{article}
\usepackage[polish,english]{babel}
\usepackage[latin1]{inputenc}
\usepackage{times}
\usepackage[T1]{fontenc}
\usepackage{epsfig}

\begin{document}

\begin{titlepage}
\vspace*{-3 cm}
\noindent

\vskip 2cm
\begin{center}
{\Large\bf 
Compact boson stars in $K$ field theories}
\vglue 1  true cm

C. Adam$^{a*}$,   N. Grandi$^{\, b**}$, P. Klimas$^{a***}$,
J. S\'anchez-Guill\'en$^{a\dagger}$,
and A. Wereszczy\'nski$^{c\dagger\dagger}$
\vspace{1 cm}

\small{ $^{a)}$Departamento de Fisica de Particulas, Universidad
     de Santiago}
     \\
     \small{ and Instituto Galego de Fisica de Altas Enerxias (IGFAE)}
     \\ \small{E-15782 Santiago de Compostela, Spain}
     \\ \small{ $^b)$ IFLP-CONICET }
     \\ \small{cc67 CP1900, La Plata, Argentina}
      \\ \small{ $^{c)}$Institute of Physics,  Jagiellonian
     University,}
     \\ \small{ Reymonta 4, 30-059 Krak\'{o}w, Poland}

\medskip
\end{center}

\normalsize
\vskip 0.2cm

\begin{abstract}
We study a scalar field theory with a non-standard kinetic term minimally coupled to gravity. We establish the existence of compact boson stars, that is, static solutions with compact support of the full system with self-gravitation taken into account. Concretely, there exist two
types of solutions, namely  compact balls on the one hand, and compact shells on the other hand. The compact balls have a naked singularity at the center. The inner boundary of the compact shells is singular, as well, but it is, at the same time, a Killing horizon. These singular, compact shells therefore resemble black holes.
\end{abstract}

\vfill

{\footnotesize
$^*$adam@fpaxp1.usc.es

$^{**}$grandi@fisica.unlp.edu.ar

$^{***}$klimas.ftg@gmail.com

$^{\dagger}$joaquin@fpaxp1.usc.es

$^{\dagger\dagger}$wereszczynski@th.if.uj.edu.pl }

\end{titlepage}

\section{Introduction}
This paper investigates self-gravitating compact solutions of a non-linear scalar field theory with a non-canonical kinetic term.
Recently it has been established that relativistic non-linear field theories may have static (solitonic) solutions with compact support, such that the fields take their vacuum values identically outside a compact region. Solitons of this type are called "compactons". At the moment, there are two known classes of field theories which may have compacton solutions. One possibility is that the (scalar) field of the theory has a potential with 
a non-continuous first derivative at (some of) its minima, a so-called V-shaped potential 
\cite{Arodz:2002yt} - \cite{kuru}.
The other possibility consists in a non-standard kinetic term (higher than second powers of the first derivatives) in the Lagrangian \cite{Adam:2007ij}, \cite{Bazeia:2008tj}, a so-called K field theory. It may be interesting to mention at this point that K field theories have found some applications already in cosmology as a candidate for dark energy, where theories of this type are known as "K essence", or "generalized dynamics", see, e.g.,  
\cite{ArmendarizPicon:1999rj} - \cite{Chimento:2003zf}. They have been also applied to topological defect formation, see, e.g., \cite{Babichev:2006cy}, \cite{Babichev:2007tn}.

In both cases of V-shaped potentials and of K theories, respectively,  the exponential approach to the vacuum typical for conventional solitons is replaced by a power-like approach. Further, the vacuum value is reached at a finite distance (the "compacton boundary"), and the second derivative of the field is non-continuous at the compacton boundary. In some cases, the compacton solutions are, therefore, weak solutions (i.e., they do not solve the field equations at the boundary) which, in these cases,  is not problematic, because the space of weak solutions is the adequate solution space for the corresponding variational problem. When gravitation is included, the stronger continuity requirements for a space-time manifold might render such a behaviour problematic, so let us emphasize already at this point that this does not happen for the self-gravitating compact solutions considered in this paper, i.e., all solutions are strong solutions of the Einstein equations.  

Like in the case of conventional solitons, for compactons it is also much simpler to find solutions in theories in 1+1 dimensions. We remark that for some non-relativistic theories (generalizations of the KdV equations) compactons were already found in 
\cite{Rosenau:1993zz},  \cite{Cooper:1993zz}. 
The most direct generalization of topological compactons to higher dimensions faces the same problems like in the case of conventional solitons, and also one possible solution is the same, namely the introduction of additional gauge fields, like in the case of vortices and monopoles 
\cite{Adam:2008rf}.  
Other possibilities to find higher-dimensional compactons consist in choosing more complicated topologies for the target space \cite{Adam:2009xb}
or in allowing for a simple time dependence in the form of Q balls 
\cite{Arodz:2008jk}, \cite{Arodz:2008nm}. 

A further natural question to be asked is whether these compactons may be coupled to gravity and whether such self-gravitating compactons lead to some interesting applications. A first application has been to brane cosmology, where the self-gravitating compactons provide a model for a thick brane with a strictly finite extension, as well as an automatic confinement of all matter fields to the brane \cite{Adam:2007ag},  \cite{Adam:2008ck}, \cite{Bazeia:2008zx}. 
As this thick brane is of the domain wall type, the resulting system  is essentially a one-dimensional problem. The coupling of the 3+1 dimensional compact Q balls of the V-shaped class of models of \cite{Arodz:2008nm} to gravity was recently studied in \cite{Kleihaus:2009kr}
mainly using numerical methods.  

It is the main purpose of the present paper to study in detail the coupling of a 3+1 dimensional radially symmetric compacton of a K field theory (with non-standard kinetic term) to gravity. 
For solutions to a scalar field plus gravity theory the notion of a "boson star" has become standard in the literature, therefore we will call generic solutions of our model "compact boson stars", and reserve more specific terms ("compact balls", or "singular shells") for the different types of solutions we shall encounter. For boson stars with a conventional kinetic term of the scalar field, there already exists a large amount of literature. Here one first criterion for a classification is whether the scalar field is real or complex. In the case of a real scalar field, a static, spherically symmetric metric requires a static, spherically symmetric scalar field. The resulting field equations are known not to support solutions when gravity is neglected, because their existence is forbidden by the Derrick theorem. For the corresponding system with gravity included, finite (asymptotic) mass solutions do exist, but they are beset by naked singularities, see e.g.   \cite{Buchdahl:1959nk}  - \cite{Xanthopoulos:1989kb}. For a complex scalar field, the Q-ball type ansatz $\phi (x) = \exp (-i\omega t) f(r)$ still is compatible with a static, spherically symmetric metric. In addition, this ansatz allows for regular, finite energy solutions already in the case without gravity. As a consequence, regular finite mass solutions  
for the full system with gravity included do exist and have been widely studied, see e.g.
\cite{Kaup:1968zz} -  \cite{Gleiser:1988rq}. Reviews on boson stars may be found, e.g., in
\cite{Jetzer:1991jr}  - \cite{Schunck:2003kk}. We remark that the system studied in the present paper is different in this respect, because, due to the non-standard kinetic term, static finite energy solutions in the non-gravitational case are not excluded by the Derrick theorem, and they do indeed exist, as we shall see in the next section.

Our paper is organized as follows.
In a first step, in Section 2 we introduce the simplest K field theory which gives rise to non-topological compact balls in flat Minkowski space (i.e., without gravity). We discuss  general features of the resulting compact ball solutions and calculate some explicit solutions by numerically integrating the ODE for spherically symmetric compact balls. In Section 3, we couple the model of the previous section to gravity via the usual, minimal coupling. We derive the Einstein equations for spherically symmetric configurations and discuss in detail their properties. We then present some explicit compact boson star solutions by numerically integrating the ODEs for the spherically symmetric configurations, and, finally, we investigate the singularities which form in these solutions. We find, in fact, two types of solutions which have two different types of singularities. The first type has naked, point-like singularities and is, thus, consistent with the "no scalar hair" conjecture. The second type has a surface-like singularity that is, at the same time, a singular boundary of space-time, and the locus of a (Killing) horizon. 
This second type of solution, therefore, exhibits some similarity with a black hole. 
Section 4 contains a discussion of these results as well as some speculations about possible astrophysical or cosmological applications of  compact boson stars of the type studied in this paper.   

\section{The non-gravitational case}

We want to study the simplest possible scalar field theory in $3+1$ dimensions with a quartic kinetic term and a standard potential which gives rise to the formation of non-topological compact solitons (i.e. compactons). Therefore we choose the Lagrangian 
\begin{eqnarray}
\mathcal{L}=-X|X|-V(\phi), 
\end{eqnarray}
where 
\begin{eqnarray}
X=\frac{1}{2}g^{\mu \nu}\partial_{\mu}\phi \partial_{\nu}\phi, \qquad V(\phi)=\lambda \phi^2.
\end{eqnarray}
We remark that, in the case at hand,  the Derrick scaling stability criterion does not exclude the existence of static solutions, thanks to the non-standard, quartic kinetic term.
The absolute value symbol in the kinetic term is necessary to ensure boundedness from below of the energy. For static configurations this absolute value symbol is immaterial in the case without gravity, because the metric functions are given functions with a fixed sign. For the case with gravitation included this is not guaranteed a priori because the metric functions have to be determined from the Einstein equations and could, in principle, change sign (as happens, e.g., in the case of the Schwarzschild solution). We will find, however, that such a sign change is not possible in our model.  

The line element $ds^2$ of flat spacetime in spherical polar coordinates reads
\begin{eqnarray}
ds^2=-dt^2+dr^2+r^2(d\theta^2+\sin^2\theta d\phi^2)
\end{eqnarray}
Now we assume spherical symmetry $\phi = \phi (r)$ then
the Euler-Lagrange equation takes the form
\begin{eqnarray}
\frac{1}{r^2}\left(r^2\phi'^3\right) '-2\lambda\phi=0.\label{ELnon}
\end{eqnarray}
It is convenient to introduce the new variable $s=r^{1/3}$  instead of $r$. In the variable $s$ equation (\ref{ELnon}) takes the form
\begin{eqnarray}
\phi'^2\phi''-54\lambda s^8\phi=0.\label{ELnon_s}
\end{eqnarray}
The advantage of the new variable $s$ is that solutions have a power series expansion about zero in the variable $s$, but not in $r$. We remark that this will no longer hold in the case with gravitation included. There solutions will have a regular power series expansion about $r=0$ in the variable $r$. For brevity, in the sequel we shall call the variable $s$ "radius", as well, and distinguish the two variables only by their respective letters.\footnote{We remark that, in addition to the compacton solutions discussed below, Eq. (\ref{ELnon_s}) has the isolated (i.e., without free integration constants) analytical solution $\phi = \pm \sqrt{\lambda /20}s^6$. This solution grows without bound for increasing $s$ and, obviously, cannot give rise to a finite energy configuration.}  
\subsection{Expansion at the boundary}
As always in the case of compactons, we now assume that there exists a certain radius $s=S$ where the field takes its vacuum value, and its first derivative vanishes.  Further, the second derivative is nonzero from below (for $s<S$), whereas it is zero for $s>S$ (such that $\phi$ takes its vacuum value zero for $s>S$). Indeed, expanding $\phi(s)$ around $s=S$,
\begin{equation} 
\phi (s) = \sum_j f_j (s-S)^j \quad {\mbox{for}} \quad s<S
\end{equation}
and assuming that $\phi(s=S)=\phi' (s=S)=0$ we get a cubic equation for $f_2$ with the three solutions 
\begin{equation} 
f_2 =(0,\pm \frac{3}{2}\sqrt{3\lambda}S^4).
\end{equation}
If we further assume, without loss of generality, that $\phi$ takes the positive value $f_2= + (3/2) \sqrt{3\lambda}S^4$ for $s<S$ and the vacuum value $f_2=0$ for $s\ge S$ then the higher coefficients $f_j$ are uniquely determined by linear equations and we find the expansion 
\begin{eqnarray}
\phi(s)&=& \frac{3}{2}\sqrt{3\lambda}S^4(s-S)^2 + \frac{12}{5}\sqrt{3\lambda}S^3(s-S)^3
+ \nonumber \\
&&2\sqrt{3\lambda}S^2(s-S)^4+\mathcal{O}((s-S)^5) \quad \mbox{for} \quad s<S \nonumber \\
\phi (s) &=& 0 \quad \mbox{for} \quad s\ge S.
\end{eqnarray}
We remark that also in the $r$ variable the function $\phi(r)$ approaches its vacuum value quadratically at $r=R\equiv S^3$, because the conditions $\phi (S)=\phi '(S)=0$ are invariant under a variable change. Indeed, we find
\begin{eqnarray}
\phi(r)&=& \frac{1}{2}\sqrt{\frac{\lambda}{3}}(r-R)^2 - \frac{1}{15R}\sqrt{\frac{\lambda}{3}}(r-R)^3 +
\nonumber \\
&& \frac{13}{270R^2}\sqrt{\frac{\lambda}{3}}(r-R)^4+\mathcal{O}((r-R)^5), \quad \mbox{for} \quad r<R \nonumber \\
\phi (r) &=& 0 \quad \mbox{for} \quad r\ge R. \nonumber
\end{eqnarray}
\subsection{Expansion at the center}
Plugging a 
Taylor series expansion around $s=0$ into eq. (\ref{ELnon_s}) we find that the first two coefficients $\phi(s=0) \equiv \phi_0$ and $\phi '(s=0) \equiv \alpha$ remain undetermined, whereas the higher ones are determined uniquely by linear equations. Due to the strong suppression factor $s^8$ at the r.h.s. of eq. (\ref{ELnon_s}) we find that the coefficients of $s^2$ to $s^9$ are, in fact, zero. Concretely we find
\begin{eqnarray}
\phi(s)=\phi_0+\alpha s+\frac{3\phi_0}{5\alpha^2}\lambda s^{10}+\frac{27}{55\alpha }\lambda s^{11}+\mathcal{O}(s^{19}).
\end{eqnarray}
The three next higher order terms $\mathcal{O}(s^{19})-\mathcal{O}(s^{21})$ include only terms proportional to $\lambda^2$. 

\subsection{Compact ball solutions}
We shall find solutions explicitly by numerical integration. But before performing these calculations we want to discuss some generic features of eq.  (\ref{ELnon_s}) which allow to establish both the existence and several properties of these compact ball solutions. A first indication for the existence of solutions can be found by counting the number of boundary conditions and free parameters in the numerical integration. There are two possibilities for the numerical integration, namely a shooting from the center, or a shooting from the boundary. If we shoot from the center, there are altogether three free parameters, namely $\phi(0)\equiv \phi_0$, $\phi'(0)\equiv \alpha$, and the compacton radius $S$. On the other hand, there are two boundary conditions that have to be obeyed for a compacton, namely $\phi (S)=\phi '(S)=0$. Therefore, we expect the existence of a one-parameter family of compact balls which may be parametrized, e.g., by the compacton radius $S$.  If we shoot from the compacton boundary, the only free parameter is the compacton radius $S$. On the other hand, there are no boundary conditions at $s=0$, because both $\phi (0)$ and $\phi '(0)$ are undetermined by the expansion at $s=0$. So again we expect a one-parameter family of solutions.

Further conclusions may be drawn by a closer inspection of eq. (\ref{ELnon_s}). Firstly, we observe that eq. (\ref{ELnon_s}) is invariant under the reflection $\phi \to -\phi$.   Therefore we may choose $\phi_0 \equiv \phi (0) >0$ without loss of generality. Secondly, it follows immediately from eq. (\ref{ELnon_s}) that away from the center (i.e. for $s>0$) it holds that if $\phi>0$ then $\phi '' >0$.
As a consequence, if $\phi_0 >0$ and $\alpha \equiv \phi '(0)>0$ then the scalar field $\phi$ will grow indefinitely and may never reach its vacuum value $\phi =0$. We conclude that a compact ball requires $\alpha <0$. Now for generic $\alpha<0$ the following may happen. If for a fixed value of $\phi_0$ the slope is too weak (i.e. if $|\alpha |$ is too small) then $\phi (s)$ will reach a point $s_0$ where $\phi '(s_0)=0$ before it reaches $\phi =0$. At this point $s_0$ the second derivative $\phi ''$ (and, consequently, all higher derivatives) becomes singular and the integration breaks down. On the other hand, if $|\alpha |$ is too big, then $\phi$ will cross the line $\phi =0$ instead of touching it, and then develop towards more and more negative values of $\phi$ indefinitely. It follows that there exists an intermediate value of $\alpha$ such that $\phi$ touches the line $\phi=0$ instead of crossing it. This is the compact ball solution, and the point $s=S$ where the touching occurs is the compacton radius. 

The explicit numerical integration confirms the behaviour described above to a high precision. In the numerical integration we mainly used shooting from the compacton boundary  
because of the lower number of free parameters, which makes the scan for compacton solutions less time consuming. A typical solution of the numerical integration is shown in Figs. 1, 2. Here, in Fig. 2 we plot the solution of Fig. 1, but using $r$ as the independent variable. The non-analytical behaviour in $r$ (the spike at $r=0$) can be clearly seen. 

Finally, let us remark that, as always in the case of compactons, we may obtain multi compacton configurations by distributing several non-overlapping compactons with different centers in flat space.   

\subsection{Energy functional}
We found that there exist compact ball solutions for arbitrary values of the field at the center $\phi_0$ (or, equivalently, for arbitrary values $S$ of the compacton radius), and it is easy to see that these different values of $\phi_0$ correspond to different values of the total energy.
Therefore, compact balls do not correspond to genuine critical points of the energy functional, and one may wonder why solutions do exist at all. 
There are two possible ways to understand this puzzle. Starting from the reduced energy functional for radially symmetric fields,
the answer is that variations of the energy functional may receive contributions from  the boundary, and it is precisely these variations of the boundary which give nonzero contributions to the energy functional for a compact ball solution. The second, equivalent answer is that the compact balls solve an inhomogeneous equation with a delta function source term.
This we want to demonstrate explicitly in the sequel. In a first step, we write down the energy functional and then use the principle of symmetric criticality to rewrite this energy functional as a functional for radially symmetric field configurations (the principle of symmetric criticality just states that this reduction of the energy functional to radially symmetric configurations provides the correct radially symmetric field equations if the field equations are compatible at all with this symmetry reduction - i.e., with the ansatz $\phi =\phi (r)$). We get
\begin{eqnarray}
E[\phi ] &=& \int r^2 dr \sin \theta d\theta d\varphi \left( \frac{1}{4} ((\nabla \phi )^2)^2 + \lambda \phi^2 \right) \nonumber \\ &=& 4\pi \int_0^\infty r^2 dr \left(\frac{1}{4} \phi_r^4 + \lambda \phi^2
\right) .
\end{eqnarray}
For the variation of the functional we find
\begin{eqnarray}
\delta E &=& 4\pi \int_0^\infty r^2 dr \left( \phi_r^3 \delta \phi_r +2\lambda \phi \delta \phi \right)
\nonumber \\
&=& 4\pi \int_0^\infty dr \left( -\partial_r (r^2 \phi_r^3 ) +2r^2 \lambda \phi \right) \delta\phi 
\nonumber \\
&&+ 4\pi (r^2\phi_r^3 \delta \phi )\vert_0^R
\end{eqnarray} 
where we performed a partial integration and used the fact that $\phi$ is a compact ball solution which takes its vacuum value for $r\ge R$. Using the small $r$ behaviour $\phi (r) \sim \phi_0 + \alpha r^\frac{1}{3} $ we finally get for the boundary term
\begin{equation}
4\pi (r^2\phi_r^3 \delta \phi )\vert_0^R = -4\pi \frac{\alpha^3}{27} \delta \phi (0) .
\end{equation}
Therefore, the variation of the energy functional for compact balls is zero only under variations which do not change the value of $\phi$ at the center, $\delta \phi (0)=0$.

A second, equivalent interpretation is like follows. We may
 require invariance of the energy functional under general variations, but then,
using $\delta \phi (0)=\int dr \delta (r) \delta \phi (r)$, we find the inhomogeneous equation with a delta function source term,
\begin{equation} \label{inhom-eq-r}
 -\partial_r (r^2 \phi_r^3 ) +2r^2 \lambda \phi -  \frac{\alpha^3}{27} \delta (r) =0 .
 \end{equation}  
Here the strength of the source term is related to the value of $\phi$ at the origin, $\phi_0$, so different $\phi_0$ (and, therefore, different energies) just correspond to different equations, i.e., different strengths of the source term. The coefficient of the source term depends directly on $\alpha \equiv \phi'(0)$, but for finite energy solutions (that is, compact balls), there exists a definite relation between $\alpha$ and $\phi_0$, so the source term determines
$\phi_0$ uniquely.  

\begin{figure}[h!]
\begin{center}
\includegraphics[width=0.65\textwidth]{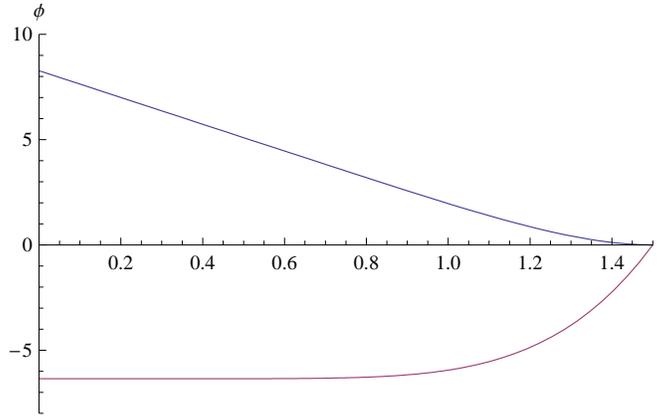}
\caption{Compacton without gravity - the profile of the scalar field $\phi(s)$ and its derivative $\phi'(s)$ for $\lambda=1$. Compacton radius $S=1.5$. Values at the center $\phi(0)=8.271$, $\phi'(0)=-6.355$.}
\end{center} 
\end{figure}

\begin{figure}[h!]
\begin{center}
\includegraphics[width=0.65\textwidth]{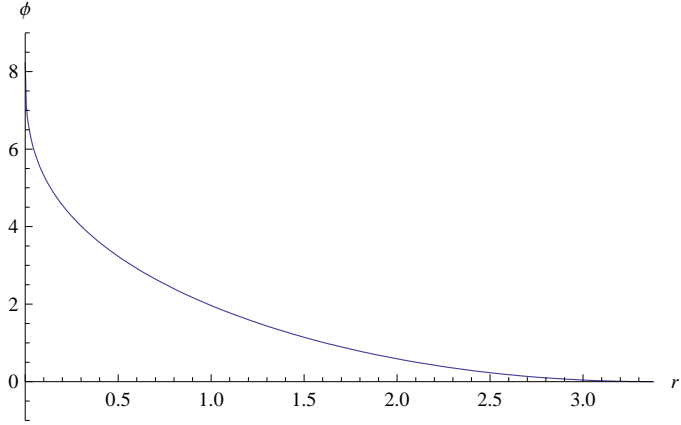}
\caption{Compacton without gravity - the profile of the scalar field of Figure 1, but expressed in the variable $r=s^3$, $\phi(r)$.  Compacton radius $R=S^3=3.375$.}
\end{center} 
\end{figure}

\section{The model with gravity}
We now consider the scalar field model of the previous section coupled minimally to gravity. The action reads
\begin{eqnarray}
\mathcal{S}=\int d^4 x \sqrt{|g|} \left( \frac{1}{\kappa^2}{\cal R}-X|X|-V(\phi) \right) , 
\end{eqnarray}
where, as before, 
\begin{eqnarray}
X=\frac{1}{2}g^{\mu \nu}\partial_{\mu}\phi \partial_{\nu}\phi, \qquad V(\phi)=\lambda \phi^2.
\end{eqnarray}
Further, ${\cal R}$ is the curvature scalar, and $g$ is the determinant of the metric tensor.
We now assume a spherically symmetric space time, then
the line element $ds^2$ may be chosen in the form
\begin{eqnarray}
ds^2=-A(r)dt^2+B(r)dr^2+r^2(d\theta^2+\sin^2\theta d\phi^2)
\end{eqnarray}
where for consistency we have to assume that $\phi = \phi (r)$ depends on the radial coordinate $r$ only.
The Euler-Lagrange equation for $\phi(r)$ takes the form
\begin{eqnarray}
\left( |B^{-1}|B^{-1}\phi'^3 \right) '+\left( \frac{2}{r}+\frac{1}{2} \left(\frac{A'}{A}+\frac{B'}{B}\right)\right)|B^{-1}|B^{-1}\phi'^3-2\lambda\phi=0 . \label{EL}
\end{eqnarray}
The Einstein equations may be obtained from the Einstein tensor, which has the independent components
\begin{eqnarray}
G_{00}&=&A\left((r B^2)^{-1}B'+\frac{1}{r^2}(1-B^{-1}) \right) , \\
G_{11}&=&B\left((rAB)^{-1}A'-\frac{1}{r^2}(1-B^{-1})\right) , \\
G_{22}&=&r^2\left(\frac{1}{2}(AB)^{-1/2}((AB)^{-1/2}A')'+\frac{1}{2}(rAB)^{-1}A'-\frac{1}{2}(rB^2)^{-1}B'\right)
\end{eqnarray}
($G_{33}$ is not independent but obeys $G_{33} = \sin^2 \theta G_{22}$), 
and the energy-momentum tensor
\begin{eqnarray}
T_{\mu \nu}&=&-\frac{\partial \mathcal{L}}{\partial (\partial^{\mu}\phi)}\partial_{\nu}\phi+g_{\mu \nu}\mathcal{L}\nonumber\\\
&=&2|X|\partial_{\mu}\phi\partial_{\nu}\phi-g_{\mu \nu}(|X|X+\lambda\phi^2)
\end{eqnarray} 
with the independent components
\begin{eqnarray}
T_{00}&=&A\left(\frac{1}{4}|B^{-1}|B^{-1}\phi'^{4}+\lambda \phi^2\right) , \\
T_{11}&=&|B^{-1}|\phi'^{4}-B\left(\frac{1}{4}|B^{-1}|B^{-1}\phi'^{4}+\lambda \phi^2\right) , \\
T_{22}&=&-r^2\left(\frac{1}{4}|B^{-1}|B^{-1}\phi'^{4}+\lambda\phi^2\right) .
\end{eqnarray}
The Einstein equations read
\begin{eqnarray}
B'&=&\frac{1}{r}B(1-B)+\kappa^2rB^2\left(\frac{1}{4}|B^{-1}|B^{-1}\phi'^4+\lambda\phi^2\right)  ,\label{eq00}\\
\frac{A'}{A}&=&\frac{1}{r}(B-1)+\frac{3}{4}\kappa^2r|B^{-1}|\phi'^4-\kappa^2\lambda rB\phi^2 , \label{eq11}
\end{eqnarray}
\begin{eqnarray}
\frac{1}{2}(AB)^{-1}A''-\frac{1}{4}(AB)^{-2}(AB)'A'+\frac{1}{2}(rAB)^{-1}A'-\frac{1}{2}(rB^2)^{-1}B'+\nonumber\\ +\kappa^2\left(\frac{1}{4}|B^{-1}|B^{-1}\phi'^{4}+\lambda\phi^2\right)
=0 . \label{eq22}
\end{eqnarray}
Here several comments are appropriate. Firstly, there seem to be four equations (the field equation (\ref{EL}) and the three Einstein equations) for three real functions $\phi$, $A$ and $B$. As always in the case of one real scalar field, these equations are, however, not independent.
The field equation may, in fact, be derived from the three Einstein equations. Further, in the case at hand, the field equation (\ref{EL}) may be derived from the three Einstein equations in a purely algebraic fashion (i.e., without performing additional derivatives). This follows easily from the fact that both the field equation (\ref{EL}) and the third Einstein equation (\ref{eq22}) are of second order. Therefore, the field equation and the third Einstein equation are completely equivalent, at least in non-vacuum regions where $\phi '\not= 0$ (in regions where $\phi '=0$ the field equation is more restrictive, because its derivation from the Einstein equations requires a division by $\phi '$). We remark that the above feature is not always true (in some cases the Einstein equations are more restrictive and have to be used, because the derivation of the field equations involves further derivatives).   

Secondly, the function $A$ appears in all equations only in the combination $(A'/A)$ and derivatives thereof (this is also true for Eq. (\ref{eq22}), where it is not completely obvious).
Therefore, one may eliminate the function $A$ with the help of Eq. (\ref{eq11}) from all the remaining equations. One may choose a set of two independent equations in $\phi$ and $B$ from these remaining equations, solve them for $\phi$ and $B$, and determine the corresponding $A$ from Eq. (\ref{eq11}) in a second step. It is obvious from Eq. (\ref{eq11}) that $A$ is determined up to a multiplicative constant. The choice of this constant just corresponds to a constant rescaling of the time coordinate. This constant will be fixed by the condition that asymptotically (that is, for $r$ bigger than the compacton radius $R$ ), the metric is equal to the Schwarzschild metric, i.e.,
\begin{equation}
A(r)=B^{-1}(r) = 1-\frac{R_s}{r} \quad \mbox{for} \quad r>R>R_s
\end{equation}  
(we will find that the compacton radius is always bigger than the Schwarzschild radius, $R>R_s$, and that a horizon never forms).
Concretely, for the two independent equations for $B$ and $\phi$ we choose Eq. (\ref{eq00}) for $B$ and the field equation Eq. (\ref{EL}) for $\phi$, where we eliminate both $(A'/A)$ and $B'$ from the latter equation with the help of Eqs. (\ref{eq11}) and (\ref{eq00}). It will be useful for our discussion to display the resulting system of two equations again. We get
\begin{equation}
\frac{B'}{B} = \frac{1}{r}(1-B)+\kappa^2 r \left( \frac{{\rm sign} (B)}{4B}\phi'^4+\lambda B\phi^2 
\right) \label{eq-B}
\end{equation}
for $B$ and
\begin{equation}
3\phi '^2 \phi '' = 2 \left( \kappa^2 \lambda r \phi^2 B - \frac{B}{r} \right) \phi'^3 + 2 \lambda {\rm sign} (B)
B^2 \phi \label{eq-phi}
\end{equation} 
for $\phi$ (here ${\rm sign}(B) \equiv (|B|/B)$ is the sign function). We remark for later use that if we forget about the sign function, then the equations for negative $B$ may be recovered from the equations for positive $B$ by the combined coupling constant transformations 
\begin{equation} \label{coupl-refl}
\lambda \to -\lambda \,  , \qquad  \kappa^2 \to - \kappa^2 . 
\end{equation}

\subsection{Qualitative behaviour of $B$}

Before starting the numerical investigation and the expansions at the compacton boundary and at the center, we want to draw some conclusions on the behaviour of the function $B$.
Concretely, we want to show that for a nonsingular scalar field $\phi$, $B(r)$ cannot approach zero from the inside, that is, from smaller values of $r$. This implies that when $B(r)$ takes the value zero at some radius $r=r_0$, then $B(r)$ is not defined for $r<r_0$. Differently stated, if we start the integration at some value $r>r_0$ where $B(r)>0$ (this we assume because we want to connect to the Schwarzschild solution), and we then integrate downwards (i.e. towards smaller values of $r$), then the integration breaks down at $r=r_0$.  
We shall find in the sequel that at this value a singularity forms, that is, some curvature invariants become infinite at $r=r_0$. Here we have to distinguish two cases, namely
$r_0=0$ or $r_0>0$. In the case $r_0 =0$ the singularity is just a point at the origin of our coordinate system, whereas for $r_0>0$ the locus of the singularity is a two-sphere $S^2$, so space has a singular boundary with $S^2$ topology. We shall call solutions of the first type "compact balls" in the sequel, whereas the second type (with the singular inner boundary of $S^2$ shape) are called "singular shells".  
 We discuss this behaviour here because it is precisely what is found in the numerical integration (that is, if we start at some compacton boundary $r=R$ and then integrate towards smaller $r$, $B$ will always hit the line $B=0$, either at $r=0$ or at some  nonzero $r=r_0$). 
  
It remains to demonstrate that $B$ cannot be continued to the region $r<r_0$. If $B>0$ for $r>r_0$ and approaches zero at $r=r_0$, then necessarily $(B'/B)>0$ for $r>r_0$ but sufficiently close to $r_0$. A continuation to $r<r_0$ may either cross zero, in which case $B<0$ and $B'>0$ for $r<r_0$ but sufficiently close to $r_0$. Or the continuation may return to the region $B>0$ for $r<r_0$, in which case $B>0$ and $B'<0$ for $r<r_0$ but sufficiently close to $r_0$. In both cases the ratio between $B'$ and $B$ is negative, $(B'/B)<0$ for $r<r_0$ but sufficiently close to $r_0$. The inequality $(B'/B)<0$ is, however, incompatible with Eq. (\ref{eq-B}) for $r<r_0$ and sufficiently close to $r_0$. In fact, the r.h.s. of Eq. (\ref{eq-B}) is manifestly positive for $r$ close to $r_0$. The first term $(1/r)(1-B) \sim (1/r)$ is positive and not small.    
The second term, $\kappa^2 r (\phi '^4/4|B|)$, is positive, and its magnitude depends on $\phi$. The third term, $\kappa^2 \lambda r B \phi^2$, may change sign if we assume that $B$ does. It is, however, small near $r_0$, because it is proportional to $B$ itself. The conclusion is that the ratio $(B'/B)$ is always positive near $r=r_0$ and, therefore, $B$ cannot be continued to values $r<r_0$.  

In a next step, we want to prove that a Schwarzschild type horizon of $B$ is impossible for compacton solutions, too. 
We remind the reader that the Schwarzschild solution for $B$ is 
\begin{eqnarray}
B_S = \left( 1-\frac{R_s}{r} \right)^{-1} = \frac{r}{r-R_s} = \frac{R_s}{\epsilon} +1
\quad , \qquad
 \epsilon \equiv r-R_s .
\end{eqnarray}
For a Schwarzschild type horizon coordinate singularity we, therefore, expand
\begin{eqnarray}
B(\epsilon ) &=& \frac{R_s}{\epsilon } + \sum_{n=0}^\infty b_n \epsilon^n \nonumber \\
\phi (\epsilon ) &=& \sum_{n=0}^\infty f_n \epsilon^n
\end{eqnarray} 
We now insert this expansion into Eq. (\ref{eq-phi}) where we assume at the moment that $r>R_s$, that is, $\epsilon>0$ and, therefore, $B>0$. Eq. (\ref{eq-phi}) then becomes
\begin{equation}
3\phi '^2 \phi '' = 2 \left( \kappa^2 \lambda (R_s + \epsilon ) \phi^2 B - \frac{B}{R_s + \epsilon } \right) \phi'^3 + 2 \lambda 
B^2 \phi \label{eq-phi-eps}
\end{equation} 
and we immediately find that $f_0 =0$ because otherwise the coefficient of $\epsilon^{-2}$ in 
Eq. (\ref{eq-phi-eps}) cannot be set to zero.
For $f_1$ we find a cubic equation with the three solutions 
\begin{equation}
f_1 = (0,\pm R_s \sqrt{\lambda} )
\end{equation}
For $f_1 =0$ it follows easily that all the higher $f_i$ are zero, as well, so that $\phi \equiv 0$ takes its vacuum value everywhere, and we are back to the Schwarzschild solution $B=B_S$ for the metric function $B$. For non-zero $f_1$ we may choose the positive root $f_1 = R_s \sqrt{\lambda}$ without loss of generality. This choice corresponds to a formal solution, but the resulting scalar field $\phi$ grows without bound for increasing $\epsilon$, as follows easily from Eq. (\ref{eq-phi-eps}). Indeed, $\phi$ can avoid unbound growth only if $\phi ' =0$ somewhere in the region $\epsilon >0$, but $\phi '=0$ implies $2\lambda B^2 \phi =0$ which is impossible ($\phi $ is nonzero at its maximum by assumption, whereas $B$ cannot approach zero in the direction of growing $r$, as we know from the last paragraph). Specifically, the formal solution for nonzero $f_1$ can, therefore, never join a compacton with its compacton boundary somewhere in the region $r>R_s$. Reversing the argument, we conclude that a horizon of the Schwarzschild type can never form for a compact ball solution. 

We still want to know what happens near a Schwarzschild type horizon for $r<R_s$, i.e., $\epsilon <0$. There $B<0$, and we remind the fact that solutions for $B<0$ may be inferred from solutions for $B>0$ by the coupling constant transformation (\ref{coupl-refl}), specifically $\lambda \to -\lambda$. We therefore find for the linear coefficient $f_1 = (0,\pm i R_s \sqrt{\lambda)}$, and obviously only the trivial vacuum solution $f_1 =0 \, \Rightarrow
\, \phi =0$ is real and, therefore, physically acceptable. We conclude that the only possible field configuration inside a horizon is the vacuum configuration $\phi \equiv 0$. We might still ask what happens if we try to put a compacton strictly inside the horizon, that is, at a compacton radius $R<R_s$. We will find that the result is the same, i.e., the nontrivial expansion coefficients for $\phi$ become imaginary, and only the trivial vacuum configuration $\phi \equiv 0$ is possible inside the horizon.      

We conclude that for compact boson star solutions the function $B$ can  never be negative, $B\ge 0$.

\subsection{Behaviour at the boundary}

Now we assume the existence of a compact boson star boundary, that is, a radius $r=R$ where
$\phi (R)=\phi '(R)=0$, whereas the second derivative is zero from above but nonzero from below.
Assuming that $B$ is nonnegative (because we want to smoothly join it to the Schwarzschild solution for $r>R$), and plugging the power series expansions around $r=R$ 
\begin{eqnarray}
\phi(r)&=&\sum_{k=2}f_k(r-R)^k,\\
B(r)&=&\sum_{k=0}b_k(r-R)^k,
\end{eqnarray}
into (\ref{eq-B}) and (\ref{eq-phi}) gives
\begin{eqnarray}
f_2&=&0, \pm \frac{1}{2}\sqrt{\frac{\lambda}{3}}b_0\nonumber\\
f_3&=&0, \mp \frac{1}{15R}\sqrt{\frac{\lambda}{3}}b_0(4b_0-3)\nonumber\\
f_4&=&0, \pm \frac{1}{270R^2}\sqrt{\frac{\lambda}{3}}b_0^2(47b_0-34)\nonumber\\
\ldots\nonumber\\
b_1&=&-\frac{b_0}{R}(b_0-1), \nonumber \\
b_2&=&\frac{b_0^2}{R^2}(b_0-1), \nonumber \\
b_3&=&-\frac{b_0^3}{R^3}(b_0-1), \nonumber \\
\ldots \nonumber
\end{eqnarray}
As expected, we find three roots for $f_2$, so the vacuum solution $f_2 =0$ for $r>R$ may be smoothly joined to one of the two nontrivial roots of $f_2$ at $r=R$. In the sequel we choose the positive root $f_2 =  (1/2)\sqrt{\lambda/3}b_0$ without loss of generality. We remark that for negative $B$ the nonzero roots of the coefficient $f_2$ become imaginary, as already announced in the previous section.
   
Inserting, further, the power series expansions above and the one for $A$,
\begin{equation}
A(r)= \sum_{k=0}^\infty a_k (r-R)^k
\end{equation}
into Eq. (\ref{eq11}) for $A$ we get
\begin{eqnarray}
 a_1 &=& \frac{a_0}{R}(b_0-1),\nonumber \\
 a_2 &=& -\frac{a_0}{R^2}(b_0-1),\nonumber \\
 a_3 &=& \frac{a_0}{R^3}(b_0-1),\nonumber \\
\ldots \nonumber
\end{eqnarray}
For the expansion coefficients of $B$ and $A$ we, therefore, find that 
also for nonzero $f_k$, the leading behaviour of $A(r)$ and $B(r)$ close to $r=R$ is like for  the vacuum (i.e., Schwarzschild) solution. The first contribution of a nonzero $f_2$ appears in $a_6$ and in $b_5$.  
An easy way to see this is to study the power series expansion of the product $AB$ for $r<R$,
\begin{eqnarray}
A(r)B(r)=a_0b_0-\frac{R}{45}\kappa^2\lambda^2a_0b_0^4(R-r)^5+\mathcal{O}((R-r)^6),
\end{eqnarray}  
where the positive nontrivial root of $f_2$ was inserted. For $r>R$, $A$ and $B$ should form the Schwarzschild metric, which fixes the constant $a_0$ to the value $a_0 = b_0^{-1}$. On the other hand, $b_0$ is a free parameter which is related to the Schwarzschild radius $R_s$ or to the asymptotic Schwarzschild mass $m_s = (R_s/2)$ of the asymptotic Schwarzschild metric, as well as to the compacton radius, via
\begin{eqnarray}
b_0=\left(1-\frac{2m_s}{R}\right)^{-1} .
\end{eqnarray}
We find that the expansion at the compacton boundary leaves us with two free parameters, namely $b_0$ and the compacton radius $R$. Equivalently, we may choose the Schwarzschild mass $m_s\equiv R_s/2$ and the compacton radius as free parameters.

Before studying the expansion  from the inside, let us discuss briefly what to expect for an integration which starts at the compacton boundary and proceeds towards smaller $r$. It holds that at $r=R$, $B(R)>1$ and $\phi (R)=\phi '(R)=0$, therefore $B'(R)$ is negative and $B$ will at first increase towards smaller values of $r$, see Eq. (\ref{eq-B}).  
For even smaller values of $r$, the additional, positive terms at the r.h.s. of Eq. (\ref{eq-B}) start to contribute, so $B'$ may become positive or negative, depending on the relative strength of the different terms at the r.h.s. of Eq. (\ref{eq-B}). It turns out numerically that for a certain radius $B'$ always becomes positive such that $B$ starts to shrink towards smaller values of $r$. Once $B$ has shrunk sufficiently such that $B<1$, then the r.h.s. of Eq. (\ref{eq-B}) is necessarily positive, such that $B$ has no other choice than shrinking further. Numerically, this is exactly what happens, for all possible values of the two free parameters. Therefore, for sufficiently small $r$, there are the following three possibilities for the behaviour of $B$. It may shrink to a nonzero value at $r=0$, i.e., $B(r=0)>0$. We will see that there exists only one isolated solution for $B(0)>0$ (i.e., without free integration constants), and this solution cannot be connected to a compacton boundary.
 Or $B$ may go to zero at $r=0$, $B(r=0)=0$. 
 We will see that two different kinds of solutions of this type may be connected to a compacton boundary. They will form the solutions of the compact ball type.
The third possibility is that $B$ becomes zero already for a nonzero $r=r_0$, i.e., $B(r=r_0>0)=0$. These are the singular shells. 

\subsection{Expansion at the center}
In this section we assume that a solution exists locally near $r=0$.  
We insert the power series expansions for $\phi$ and $B$
\begin{eqnarray}
\phi(r)&=&\sum_{k=0}\bar f_k r^k,\\
B(r)&=&\sum_{k=0}\bar b_k r^k,
\end{eqnarray}
into (\ref{eq-B}) and (\ref{eq-phi}). In a first step we want to assume that $\bar b_0 >0$. Cancellation of the coefficient of $r^{-1}$ in Eq. (\ref{eq-B}) then requires $\bar b_0 =1$, whereas cancellation of the coefficients of $r^{-1}$ and $r^0$ in Eq. (\ref{eq-phi}) require $\bar f_0 =0$ and $\bar f_1 =0$. All higher expansion coefficients are uniquely determined (up to an overall sign of $\phi$), that is, the solution is an isolated one without free integration constants. Explicitly, we find (we choose the plus sign for $\phi$)   
\begin{eqnarray}
\phi(r)&=& \frac{\sqrt{\lambda}}{\sqrt{20}} r^2  
+\mathcal{O}(r^{8}) \nonumber \\
B(s)&=& 1 +  \frac{3\kappa^2 \lambda^2}{350}  r^6 +
\mathcal{O}(r^{12})
\end{eqnarray}
This solution behaves like at a compacton boundary already at the center $r=0$ (i.e. $\phi (0)=0$, $\phi '(0)=0$), and the question is whether it can be connected to a compacton boundary at some nonzero $r=R$. It follows easily from Eq. (\ref{eq-phi}) that this is impossible. Indeed, if $\phi$ takes its vacuum value $\phi =0$ at two different radii, then it must pass through a local maximum at some $r=r_0$ between the two radii (where $\phi (r_0)>0$ and $\phi '(r_0)=0$).
But the existence of this maximum is incompatible with Eq. (\ref{eq-phi}).

Therefore, we now assume $\bar b_0=0$ and find
\begin{eqnarray}
\phi(r)&=& \bar f_0+\bar f_1 r -\frac{1}{3}\bar b_1 \bar f_1 r^2  + \left( \frac{5}{27}\bar b_1^2 \bar f_1 -\frac{1}{36}\kappa^2 \bar f_1^5 \right) r^3 
+\mathcal{O}(r^{4}) \nonumber \\
B(s)&=& \bar b_1 r + \left( \frac{1}{4} \kappa^2 \bar f_1^4 - \bar b_1^2 \right) r^2 +
\left( \bar b_1^3-\frac{7}{12} \kappa^2 \bar b_1 \bar f_1^4 \right) r^{3}+
\mathcal{O}(r^{4}) .
\end{eqnarray}
Here, $\bar f_0$, $\bar f_1$ and $\bar b_1$ are free parameters (integration constants).
We remark that for compact ball solutions (that is, for the correct matching to a compacton boundary) both $\bar f_0$ and $\bar f_1$ have to be nonzero. On the other hand, there will exist solutions with $\bar b_1=0$. In this case the leading terms of the expansion read
\begin{eqnarray}
\phi(r)&=&\bar f_0+\bar f_1 r -\frac{\kappa^2}{36} \bar f_1^5 r^3+\frac{\kappa^4 \bar f_1^5 (5\bar f_1^4+18\lambda\bar f_0^2)}{2160}r^{5}+\mathcal{O}(s^{6})\nonumber \\
B(s)&=&\frac{\kappa^2}{4}\bar f_1^4 r^2-\frac{7\kappa^4 \bar f_1^8}{144}r^{4}+\frac{\kappa^6 \bar f_1^8(4\bar f_1^4+9\lambda\bar f_0^2)}{432}r^{6}+\mathcal{O}(r^{7})
\end{eqnarray}
The power series expansion for $A$ has to be treated independently for $\bar b_1 \not= 0$ and $\bar b_1 =0$, because the leading behaviour for small $r$ is completely different, being $A(r) \sim r^{-1}$ for $\bar b_1 \not= 0$, and $A \sim r^2$ for $\bar b_1 =0$, as may be inferred easily from Eq. (\ref{eq11}). For $\bar b_1 \not= 0$
the leading terms in the expansion read
\begin{eqnarray}
A(r) &=& \bar a_{-1} \left( \frac{1}{r} +  \left( b_1 + \frac{3\kappa^2 \bar f_1^4}{4 \bar b_1} \right) + \frac{\kappa^2 \bar f_1^4}{16 \bar b_1^2} \left( 4\bar b_1^2 + 3 \kappa^2 \bar f_1^4 \right) r + \mathcal{O} (r^2) \right) 
\end{eqnarray} 
whereas for $\bar b_1 =0$ it reads
\begin{eqnarray}
A(r)&=&\bar a_2 \left(  r^2-\frac{ \kappa^2 \bar f_1^4}{12}r^4+\frac{ \kappa^4 \bar f_1^8 }{108}r^6 +\mathcal{O}(r^{8})
\right) 
\end{eqnarray}
The leading coefficients $\bar a_{-1}$ or $\bar a_2$, respectively, are free parameters in the power series expansion. They appear as linear factors also in all higher coefficients, in accordance with our observation that $A$ is determined up to a multiplicative constant. They cannot be determined from the local analysis at $r=0$, but must instead be determined from the condition that $A$ approaches a Schwarzschild metric at the compacton boundary.

\subsection{Selfgravitating compact ball solutions}

Before explicitly performing the numerical integration, we again want to determine the number of free parameters and boundary conditions, both for a shooting from the boundary and for a shooting from the center. Here we only consider the parameters and boundary conditions for $\phi$ and $B$, because these two can be determined from the system of two equations 
(\ref{eq-B}) and (\ref{eq-phi}). $A$ may then be determined from Eq. (\ref{eq11}) in a second step, and we know already that the multiplicative constant which is the free parameter of $A$ must be determined by a matching to the Schwarzschild metric at the compacton boundary.

At the compacton boundary there are two free parameters, namely the compacton radius $R$ and the Schwarzschild (or asymptotic, ADM) mass $m_s$. Concerning the conditions that have to be imposed at the center $r=0$, we have to distinguish the case $\bar b_1 \not= 0$ from the case $\bar b_1 =0$. In the case $\bar b_1 \not= 0$, there are no conditions imposed at $r=0$, because the first two coefficients $\bar f_0$ and $\bar f_1$ of $\phi$ are unrestricted, and the fact that $\bar b_0 =0$ does not count as a boundary condition, because it is true for all generic solutions that exist locally near $r=0$ (we remind the reader that the solution with $\bar b_0 =1$ is an isolated solution with no free parameters). We, therefore, expect a two-parameter family of solutions. Specifically, for a fixed Schwarzschild mass we expect to find a one-parameter family of compact ball solutions with different radii. We shall call this type of solutions with $\bar b_1>0 $ "large compactons" in the sequel.

In the case $\bar b_1 =0$, this condition provides exactly one boundary condition at $r=0$. In this case we, therefore, expect a one-parameter family of solutions. Specifically, for a fixed Schwarzschild mass $m_s$ we expect only one compact ball with a fixed radius.  We shall refer to this type of solutions as "small compactons".

An analysis of the shooting from the center leads to the same results. For the case $\bar b_1 \not= 0$ (large compactons), there are four free parameters, namely $\bar b_1$ itself, $\bar f_0$, $\bar f_1$, and the compacton radius $R$. Further, there are two boundary conditions at the compacton boundary $r=R$, namely the conditions $\phi (R)=0$ and $\phi ' (R) =0$. Therefore, we expect a two-parameter family of compact ball solutions. In the case $\bar b_1 =0$ (small compactons), we are left with three free parameters $\bar f_0$, $ \bar f_1$ and $R$ and the same boundary conditions at $r=R$, therefore we expect a one-parameter family of solutions.

The explicit numerical integration completely confirms the above results. Concretely, we prefer to shoot from the boundary because of the smaller number of free parameters. There, the free parameters are the Schwarzschild mass $m_s = R_s/2$ (where $R_s$ is the Schwarzschild radius), and the compacton radius $R$. In the numerical integration we choose a fixed $m_s$ and then perform the integration for different values of $R$. We already know from the discussion of the previous sections that a solution cannot exist for $R\le R_s$. For $R>R_s$, we find the following behaviour. If $R$ is too small, then in the integration from $r=R$ towards the center $B$ reaches zero already at a nonzero radius $r=r_0$, i.e., $B(r_0)=0$. 
These singular shell solutions shall be discussed in the next subsection.
Increasing $R$ further, it reaches a minimum value $R=R_{sc}$ (where $sc$ stands for "small compacton") such that $B$ approaches zero at $r=0$. For this minimum value of $R$, $B$ approaches zero quadratically, $B\sim r^2$, i.e., it corresponds to the case $\bar b_1 =0$, the small compacton. Two examples (for $m_s =0.1$ and $m_s =1.5$, respectively), are shown in Figures 3-8. 

\begin{figure}[h!]
\begin{center}
\includegraphics[width=0.65\textwidth]{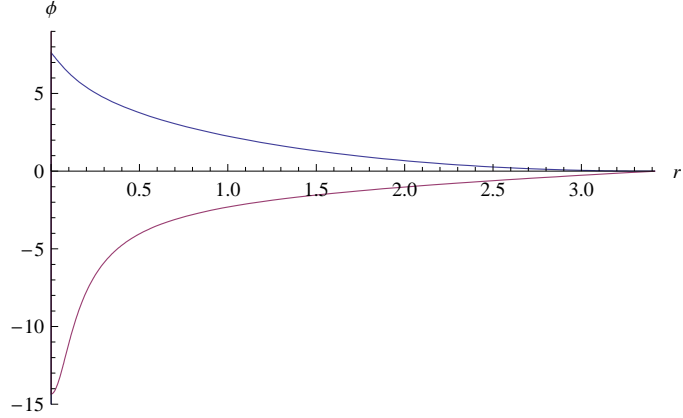}
\caption{Compacton with gravity included, small compacton case ($\bar b_1 =0$). Profile of the scalar field and its derivative for $\lambda=1$ and $\kappa=0.1$. Shooting from the boundary: initial value $m_s=0.1$, resulting compacton radius $R_{sc}=3.415$. Values at the center $\phi(0)=7.633$ and $\phi'(0)=-14.341$.}
\end{center} 
\end{figure}

\begin{figure}[h!]
\begin{center}
\includegraphics[width=0.65\textwidth]{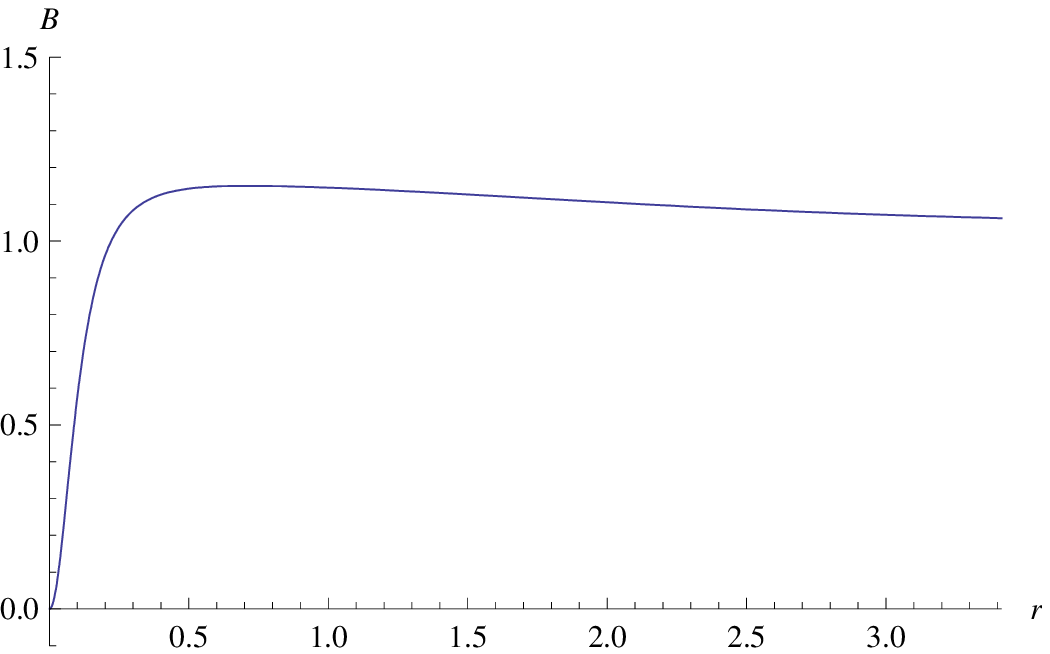}
\caption{Small compacton case ($\bar b_1 =0$). Function $B(r)$ for $\lambda=1$ and $\kappa=0.1$. Shooting from the boundary, like in Figure 3: $m_s=0.1$, $R_{sc}=3.415$.}
\end{center}
\end{figure} 

\begin{figure}[h!]
\begin{center}
\includegraphics[width=0.65\textwidth]{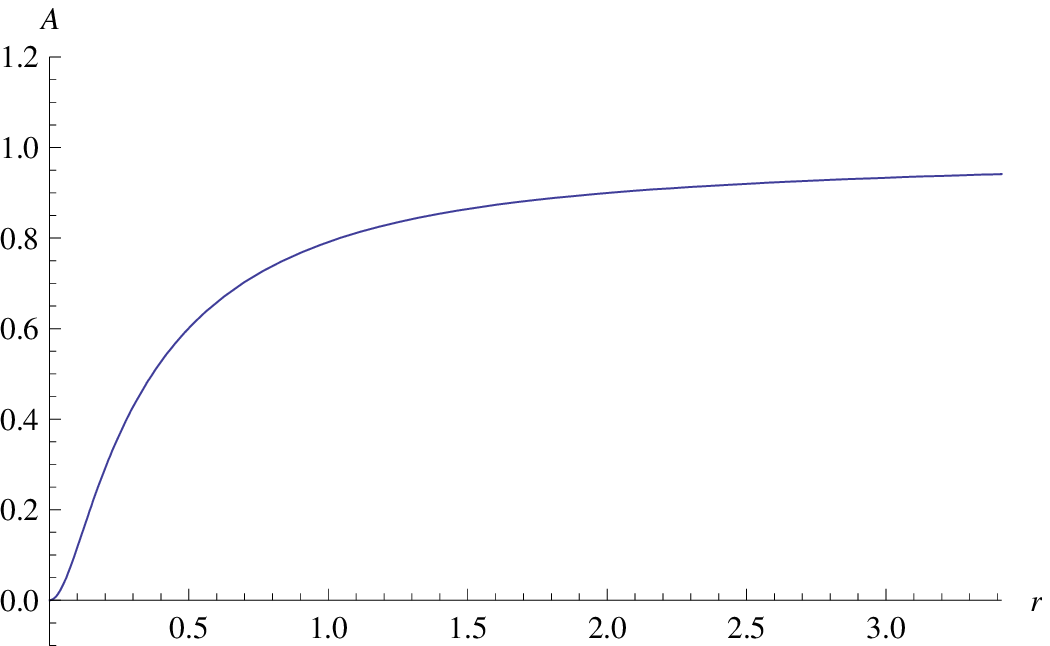}
\caption{Small compacton case ($\bar b_1 =0$). Function $A(r)$ for $\lambda=1$ and $\kappa=0.1$. Shooting from the boundary, like in Figure 3: $m_s=0.1$, $R_{sc}=3.415$. }
\end{center} 
\end{figure}

\begin{figure}[h!]
\begin{center}
\includegraphics[width=0.65\textwidth]{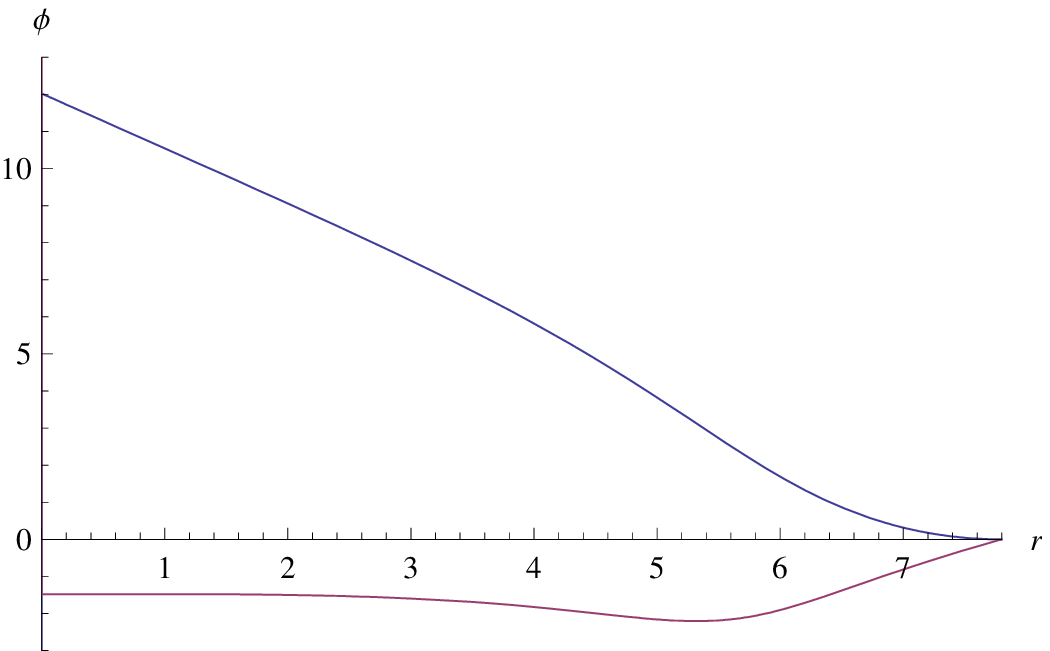}
\caption{Compacton with gravity included, small compacton case ($\bar b_1 =0$).  Profile of the scalar field and its derivative for $\lambda=1$ and $\kappa=0.1$. Shooting from the boundary: initial value $m_s=1.5$, resulting compacton radius $R_{sc}=7.800$.  Values at the center $\phi(0)=12.021$ and $\phi'(0)=-1.481$.}
\end{center} 
\end{figure}

\begin{figure}[h!]
\begin{center}
\includegraphics[width=0.65\textwidth]{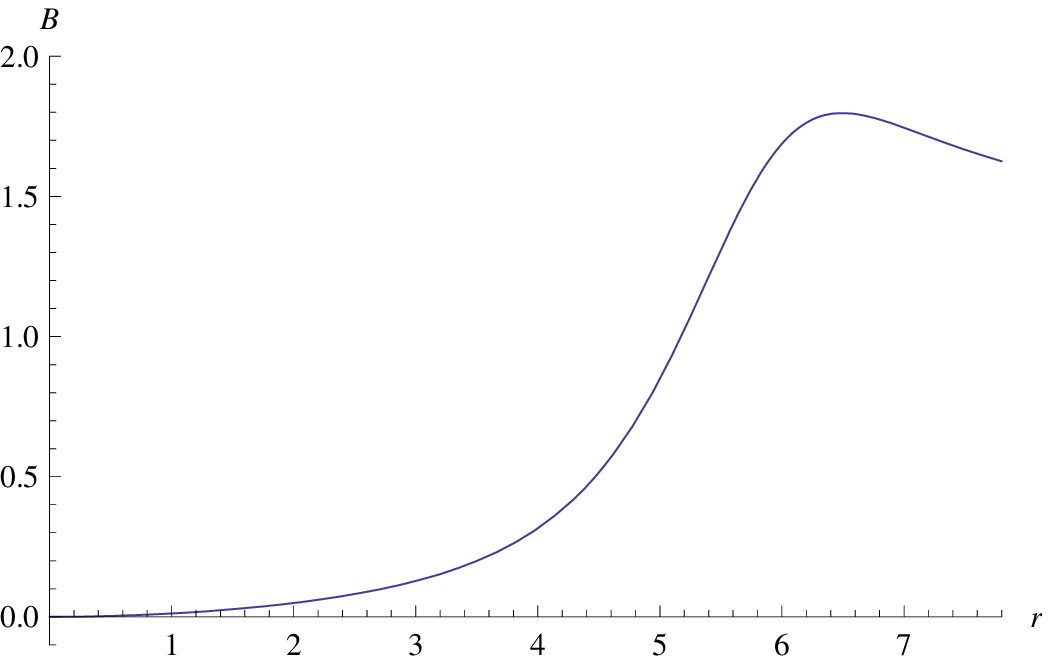}
\caption{Small compacton case ($\bar b_1 =0$). Function $B(r)$ for $\lambda=1$ and $\kappa=0.1$. Shooting from the boundary, like in Figure 6: $m_s=1.5$, $R_{sc}=7.800$.}
\end{center}
\end{figure} 

\begin{figure}[h!]
\begin{center}
\includegraphics[width=0.65\textwidth]{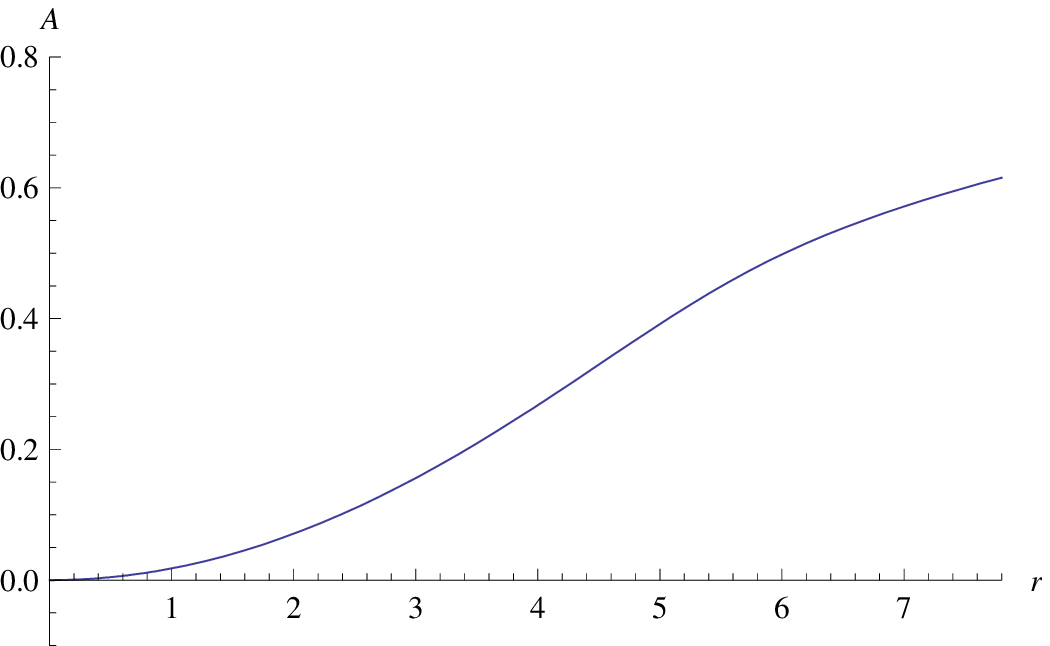}
\caption{Small compacton case ($\bar b_1 =0$). Function $A(r)$ for $\lambda=1$ and $\kappa=0.1$. Shooting from the boundary, like in Figure 6: $m_s=1.5$, $R_{sc}=7.800$.}
\end{center} 
\end{figure}

Further, we show, in Figures 9-11
the dependence of $R$ and the values of $\phi $ and $\phi '$ at the origin $r=0$ as a function of the Schwarzschild mass $m_s$ (which we use as the only free parameter in the case of small compactons). We remark that for sufficiently large Schwarzschild mass $m_s$, the relation between the compacton radius $R_{sc}Ð$ and the Schwarzschild mass $m_s$ plotted in Fig. 9 may be well approximated by the linear interpolation formula
\begin{equation}
R_{sc} \sim 2.9 m_s + 3.4 >2m_s ,
\end{equation}
and $R_{sc}$ is always above the Schwarzschild radius $R_s = 2m_s$, i.e., $R_{sc}>R_s$ (this inequality remains true also for small values of $m_s$ where the above linear interpolation cannot be used).

\begin{figure}[h!]
\begin{center}
\includegraphics[width=0.65\textwidth]{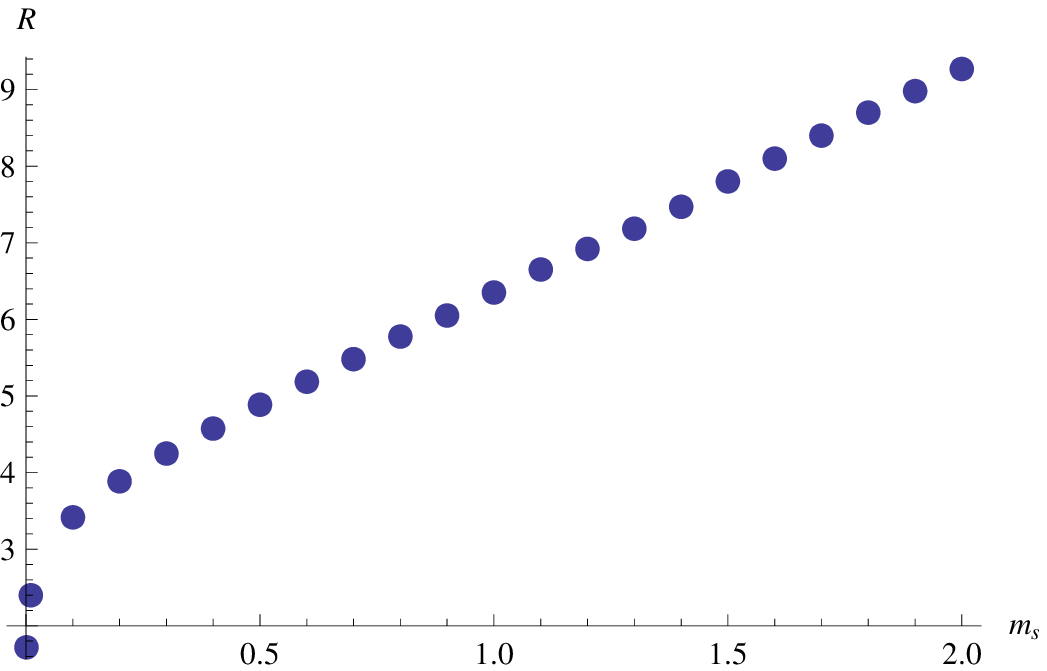}
\caption{Small compacton case ($\bar b_1=0$). Radius $R_{sc}$ of the small compacton as a function of the Schwarzschild mass $m_s$.}
\end{center}
\end{figure} 

\begin{figure}[h!]
\begin{center}
\includegraphics[width=0.65\textwidth]{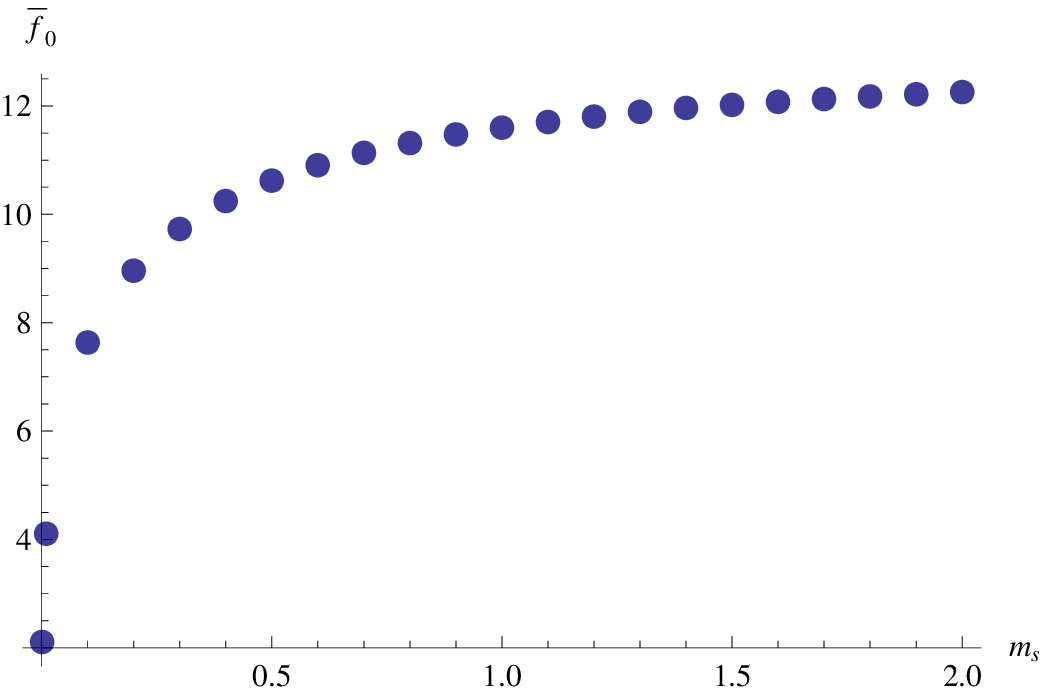}
\caption{Small compacton case ($\bar b_1=0$). Value of the scalar field at at the center, as a function of the Schwarzschild mass, $\overline{f}_0(m_s)$.}
\end{center}
\end{figure} 

\begin{figure}[h!]
\begin{center}
\includegraphics[width=0.65\textwidth]{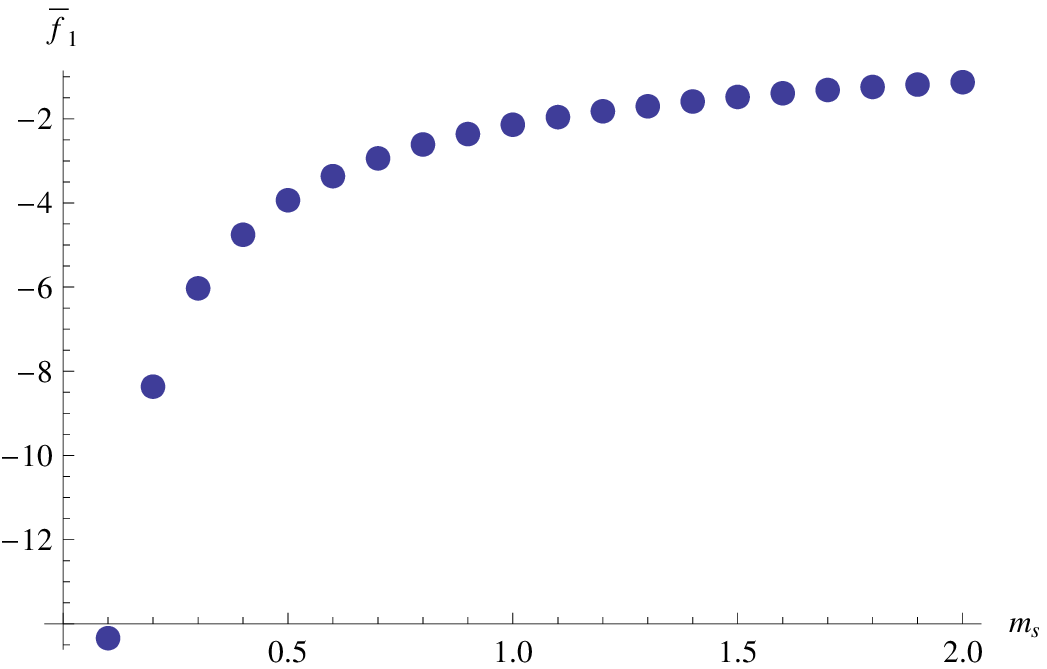}
\caption{Small compacton case ($\bar b_1=0$). Value of the first derivative of the scalar field at at the center, as a function of the Schwarzschild mass, $\overline{f}_1(m_s)$.}
\end{center}
\end{figure}

 For values of $R$ which are larger than $R_{sc}$ for a given $m_s$ (the radius of the small compacton), $B$ approaches zero linearly. This corresponds to the case $\bar b_1 >0$, i.e., to the large compacton. There exist solutions of this type with arbitrarily large compacton radius $R$ for one and the same value of the Schwarzschild mass. Some examples are shown in Figures 12-14 for the fixed value $m_s=1.5$ and three different values of the compacton radius.
It can be seen from Figures 13, 14, that the functions $A$ and $B$ approach the same Schwarzschild solution (with the same Schwarzschild mass $m_s =1.5$) for sufficiently large radius. There exist, in fact, some subtleties related to the large compacton solutions, which we want to discuss now. First of all, smaller values of $\bar b_1$ do {\em not} correspond to smaller values of the compacton radius (specifically, the radius of the small compacton is not reached by the limit $\bar b_1 \to 0$ of the large compacton radius). Quite on the contrary, small $\bar b_1$ corresponds to large compacton radius, as can be seen easily in Figure 13. This demonstrates again that the small compacton case $\bar b_1 =0$ cannot be seen as the limiting case $\lim \bar b_1 \to 0$ of the large compacton case. 

\begin{figure}[h!]
\begin{center}
\includegraphics[width=0.65\textwidth]{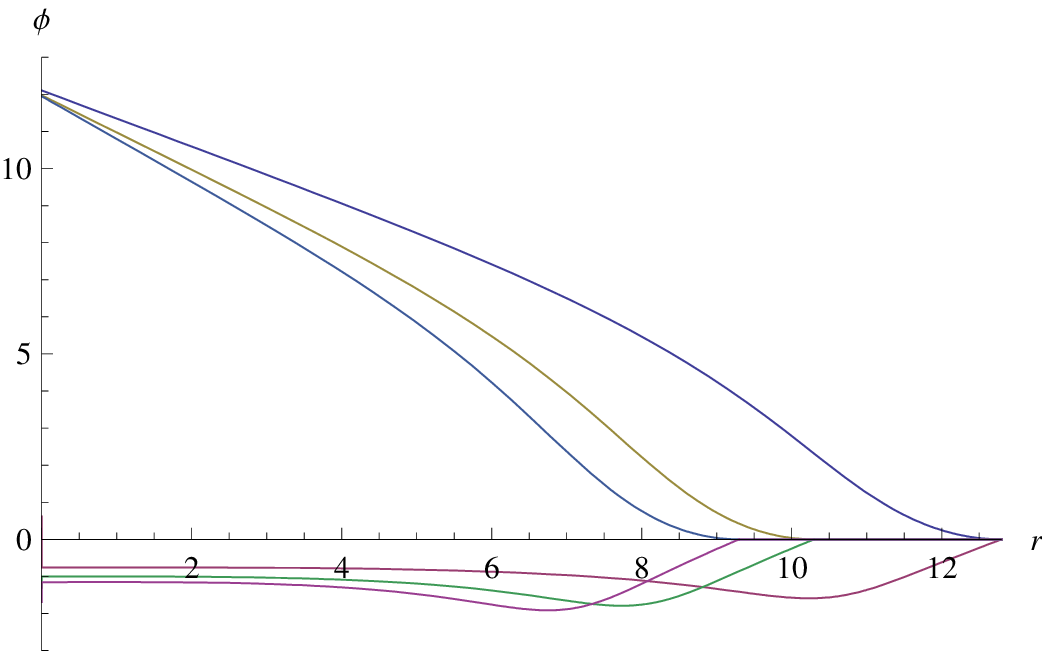}
\caption{Compacton with gravity included, large compacton case ($\bar b_1 >0$). Profile of the scalar field and its derivative for $\lambda=1$ and $\kappa=0.1$. Shooting from the boundary: initial value     $m_s=1.5$, and the three initial values for the compacton radius $R=9.3$, $R=10.3$ and $R=12.8$.}
\end{center} 
\end{figure}

\begin{figure}[h!]
\begin{center}
\includegraphics[width=0.65\textwidth]{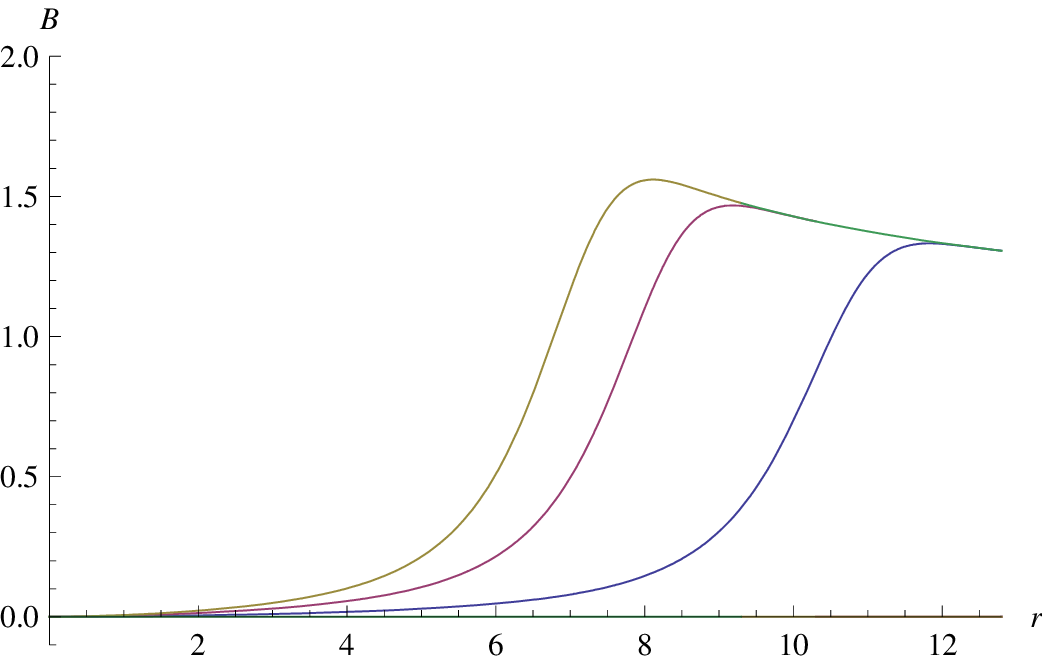}
\caption{Large compacton case ($\bar b_1 >0$). Function $B(r)$ for $\lambda=1$ and $\kappa=0.1$. Shooting from the boundary, like in Figure 12: $m_s=1.5$, and $R=9.3$, $R=10.3$ and $R=12.8$.}
\end{center}
\end{figure} 

\begin{figure}[h!]
\begin{center}
\includegraphics[width=0.65\textwidth]{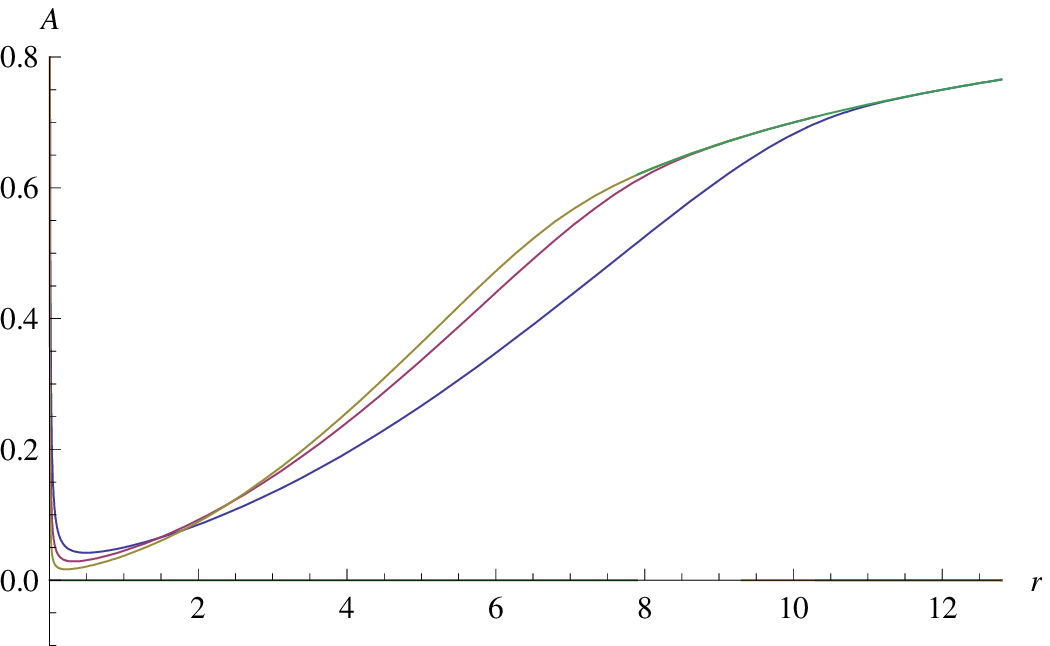}
\caption{Large compacton case ($\bar b_1 >0$). Function  $A(r)$ for $\lambda=1$ and $\kappa=0.1$. Shooting from the boundary, like in Figure 12:  $m_s=1.5$, and $R=9.3$, $R=10.3$ and $R=12.8$.}
\end{center} 
\end{figure}

A second subtlety can be seen once we reverse the role of $m_s$ and $R$ in the case of the large compactons. That is to say, we choose now a fixed $R$ and then vary $m_s$. For too large $m_s$, no solution exists. For a certain limiting value of $m_s$, we get the small compacton, and for even smaller $m_s$, we find the large compactons. The question arises how small $m_s$ can get for a given $R$ such that a large compacton exists. The answer is that it can get arbitrarily small. Specifically, there exist large compactons for negative values of the Schwarzschild mass $m_s$.  We show an example of this fact in Figures 15-17.

\begin{figure}[h!]
\begin{center}
\includegraphics[width=0.65\textwidth]{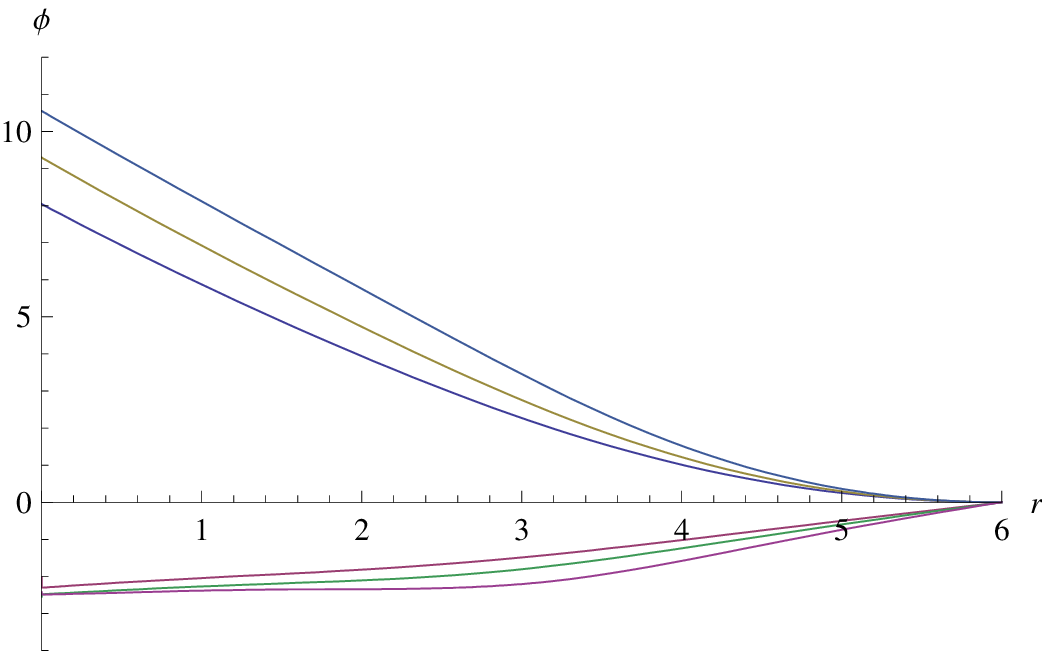}
\caption{Large compacton case ($\bar b_1 >0$). Profile of the scalar field and its derivative for $\lambda =1$ and $\kappa = 0.1$. Shooting from the boundary for the initial value $R=6$ and the three initial values
$m_s = 0.5 $ (upper curve), $m_s =0$ (middle curve) and $m_s =-0.5$ (lower curve). }
\end{center} 
\end{figure}

\begin{figure}[h!]
\begin{center}
\includegraphics[width=0.65\textwidth]{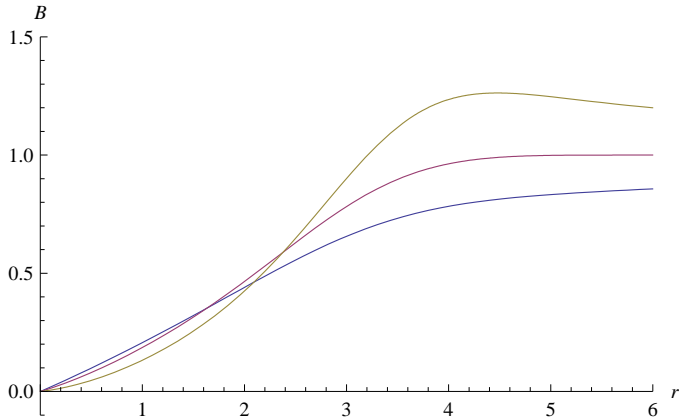}
\caption{Large compacton case ($\bar b_1 >0$). Function $B(r)$ for $\lambda =1$ and $\kappa = 0.1$. Shooting from the boundary for the initial value $R=6$ and the three initial values
$m_s = 0.5 $, $m_s =0$ and $m_s =-0.5$. The graph for positive $m_s$ has a maximum and decreases for larger $r$, the graph for $m_s=0$ approaches the constant value 1 at $r=R$, whereas the graph for negative $m_s$ never reaches $B=1$ and has positive slope for all $r$. }
\end{center} 
\end{figure}

\begin{figure}[h!]
\begin{center}
\includegraphics[width=0.65\textwidth]{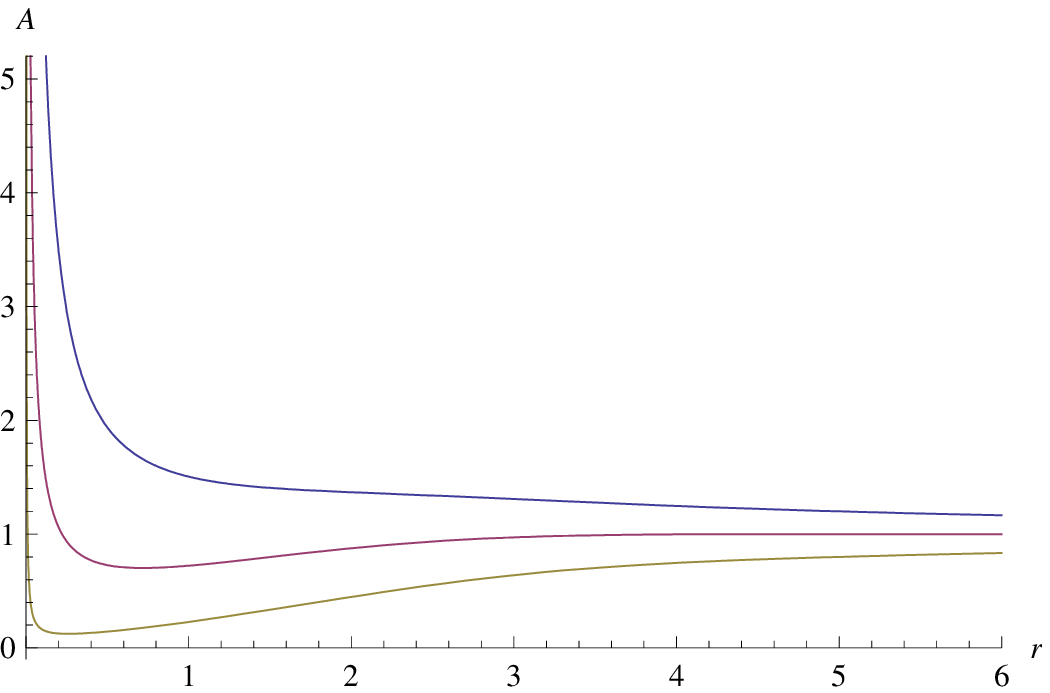}
\caption{Large compacton case ($\bar b_1 >0$). Function $A(r)$ for $\lambda =1$ and $\kappa = 0.1$. Shooting from the boundary for the initial value $R=6$ and the three initial values
$m_s = 0.5 $ (lower curve), $m_s =0$ (middle curve), and $m_s =-0.5$ (upper curve).  }
\end{center} 
\end{figure}

The existence of large compacton solutions for a negative Schwarzschild mass may be further understood by the introduction of a "variable mass" function $m(r)$ as a new variable instead of $B$,
\begin{equation}
B(r) \equiv \left( 1-\frac{2m (r)}{r} \right)^{-1} = \frac{r}{r-2m(r)} .
\end{equation}
The condition that $B(r=0)=0$ just requires that $m(r)$ should not go to zero too fast in the limit $r\to 0$ ($m(r=0)\not= 0$ is a sufficient condition), whereas the condition $B' (r=0)=\bar b_1 >0$ leads to
\begin{equation}
m_0 \equiv m(r=0)= -\frac{1}{2\bar b_1} <0.
\end{equation} 
This relation will explain the unexpected relation between $\bar b_1$ and the radius $R$ of the large compacton. In addition, this relation demonstrates, once again, that the small compacton case cannot be obtained as the limit of the large compacton case (the function $m(r)$ is, in fact, not useful for the small compacton, because it does not have a power series expansion at $r=0$ for the small compacton case). Inserting $m(r)$ into Eq. (\ref{eq-B}) for $B$ we get  
\begin{equation}
m' = \frac{\kappa^2}{2} \left( (r-2m)^2 \phi '^4 + \lambda r^2 \phi^2 \right)
\end{equation}
with the formal solution
\begin{equation}
m(r) = \int_0^r d\tilde r  \frac{\kappa^2}{2} \left( (\tilde r-2m)^2 \phi '^4 + \lambda \tilde r^2 \phi^2 \right) + m_0
\equiv \tilde m(r) + m_0
\end{equation}
and, therefore, for the Schwarzschild mass
\begin{equation}
m_s \equiv m(r=R) = \tilde m(R) + m_0.
\end{equation}
 We observe that the integrand in the definition of $\tilde m(r)$ is non-singular and positive definite, therefore $\tilde m(R)$ is a monotonously increasing function of $R$ and obeys $\tilde m(R=0)=0$. Now we treat $R$ and $m_0$ (that is, $R$ and $\bar b_1$) as our two free parameters for the large compacton solutions. The only restrictions on these parameters are $R>0$ and $m_0<0$ (that is, $\bar b_1 >0$). Obviously, for any choice of $R>0$ there always exists a choice for $m_0<0$ such that  the Schwarzschild mass is negative, $m_s <0$. One may, of course, exclude such choices as unphysical. We just want to remark that from the point of view of the large compacton solutions there is nothing special about negative $m_s$, i.e., solutions with negative $m_s$ may be reached from positive $m_s$ by a completely smooth variation of the parameters which characterize these solutions.     
  
\subsection{Singular shells}
Now we want to discuss the case of singular shells, that is, of solutions which connect a 
singular inner boundary at $r=r_0$ with a compacton boundary at $r=R$. Here, the singular boundary is defined by $B(r_0)=0$, and it holds that $r_0<R$.   In a first step, we  insert the power series expansions about $r=r_0$
\begin{eqnarray}
\phi(r)&=&\sum_{k=0}\bar f_k (r - r_0)^k,\\
B(r)&=&\sum_{k=1}\bar b_k (r - r_0)^k,
\end{eqnarray}
into (\ref{eq-B}) and (\ref{eq-phi}), where we assume $B(r=r_0)=0$. We get
\begin{eqnarray}
\phi(r)&=& \bar f_0+\bar f_1 (r-r_0)   + \frac{\kappa^2 \bar f_1^5}{36}\left( \kappa^2 \lambda r_0^2 \bar f_0^2 -1 \right) (r - r_0)^3 
+\mathcal{O}((r - r_0)^{4}) \nonumber \\
B(s)&=&  \frac{1}{4} \kappa^2 r_0 \bar f_1^4 (r - r_0) +  \frac{1}{4} \kappa^2 \bar f_1^4  (r - r_0)^2 + \nonumber \\ &&
\frac{7}{144} \kappa^4 r_0 \bar f_1^8 \left( \kappa^2 \lambda r_0^2 \bar f_0^2 \right) (r - r_0)^{3}+
\mathcal{O}((r - r_0)^{4}) .
\end{eqnarray}
Here, $\bar f_0$ and $\bar f_1$ are free parameters (integration constants), whereas $\bar b_1$ is already determined by the equations. Inserting the corresponding power series expansion for $A$ into Eq. (\ref{eq11}), we find that the first nonzero coefficient is the cubic one in this case. Explicitly, we get
\begin{eqnarray}
A(r)&=&\bar a_3 \left(  (r - r_0)^3-\frac{ 1}{r_0}(r - r_0)^4+  \right. \nonumber \\ && \left. 
\left( \frac{1}{r_0^2} + \frac{ \kappa^2 \bar f_1^4 }{12} \left( \kappa^2 \lambda r_0^2 \bar f_0^2 -1 \right)  \right) (r- r_0)^5 +\mathcal{O}((r - r_0)^{6}) \right) .
\end{eqnarray}
Next, let us count free parameters and conditions in this case. For a shooting from the outer compacton boundary, we have three free parameters, namely $m_s$, $R$ and $r_0$. On the other hand, we have one condition, namely $B(r_0)=0$. So we expect a two-parameter family of solutions. If we shoot from the inner boundary, we have the four free parameters $\bar f_0$, $\bar f_1$, $R$ and $r_0$, and the two conditions $\phi (R) = \phi '(R) =0$. Again, we expect a two-parameter family of solutions. Specifically, for a fixed asymptotic mass $m_s$ we expect a one-parameter family of solutions which may be parametrized, e.g., by the compacton radius $R$. This is in complete accordance with the discussion of the previous section where we found already that for fixed $R_s \equiv 2 m_s$ there should exist singular shell solutions for $R_s < R < R_{sc}$ (here $R_{sc}$ is the radius of the small compacton). It turns out that the inner singular radius of the singular shell is always smaller than the Schwarzschild radius, so we have, in fact, $r_0 < R_s < R$ for all singular shells.

\begin{figure}[h!]
\begin{center}
\includegraphics[width=0.65\textwidth]{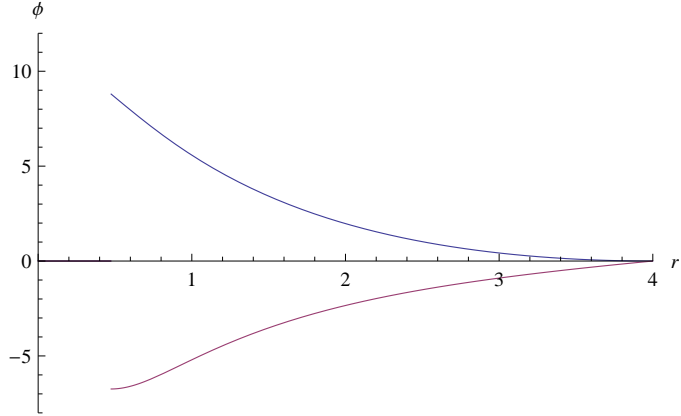}
\caption{Singular shell case. Functions $\phi(r)$ and $\phi '(r)$ for $\lambda =1$ and $\kappa = 0.1$. Shooting from the compacton boundary for the initial values $m_s = 0.5$ and $R=4$. The resulting singular boundary is at $r_0 = 0.4736$ (position of the minimum of $\phi '$). }
\end{center} 
\end{figure}

\begin{figure}[h!]
\begin{center}
\includegraphics[width=0.65\textwidth]{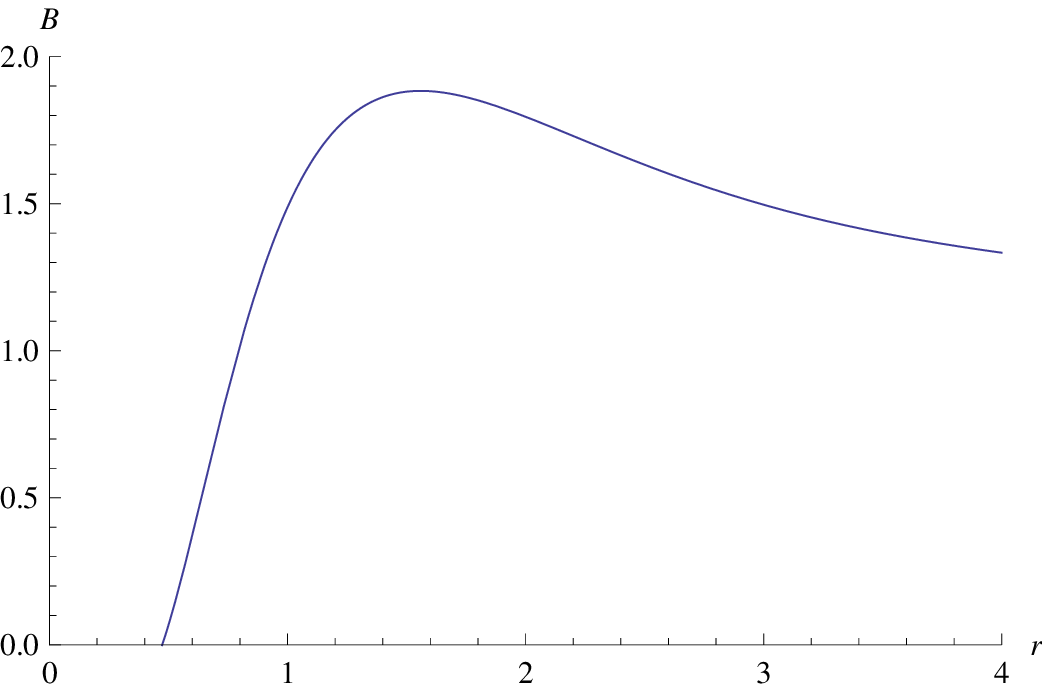}
\caption{Singular shell case. Function $B(r)$ for $\lambda =1$ and $\kappa = 0.1$. Shooting from the compacton boundary for the initial values $m_s = 0.5$ and $R=4$. The resulting singular boundary is at $r_0 = 0.4736$ (position of the zero of $B$).   }
\end{center} 
\end{figure}

\begin{figure}[h!]
\begin{center}
\includegraphics[width=0.65\textwidth]{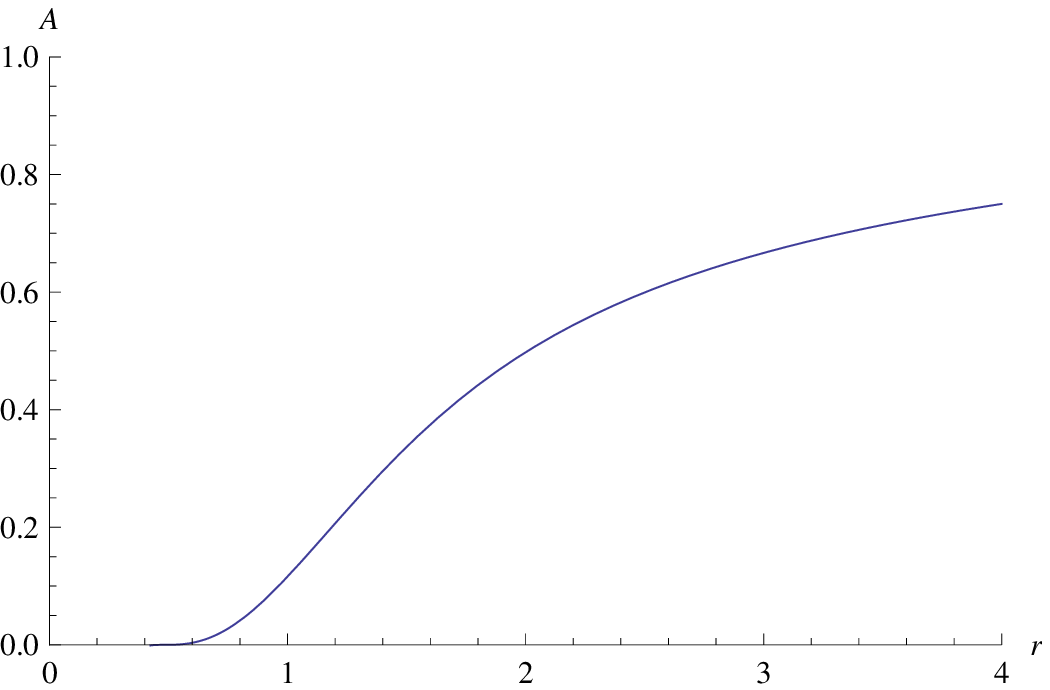}
\caption{Singular shell case. Function $A(r)$ for $\lambda =1$ and $\kappa = 0.1$. Shooting from the compacton boundary for the initial values $m_s = 0.5$ and $R=4$. The resulting singular boundary is at $r_0 = 0.4736$ (position of the zero of $A$).   }
\end{center} 
\end{figure}

\begin{figure}[h!]
\begin{center}
\includegraphics[width=0.65\textwidth]{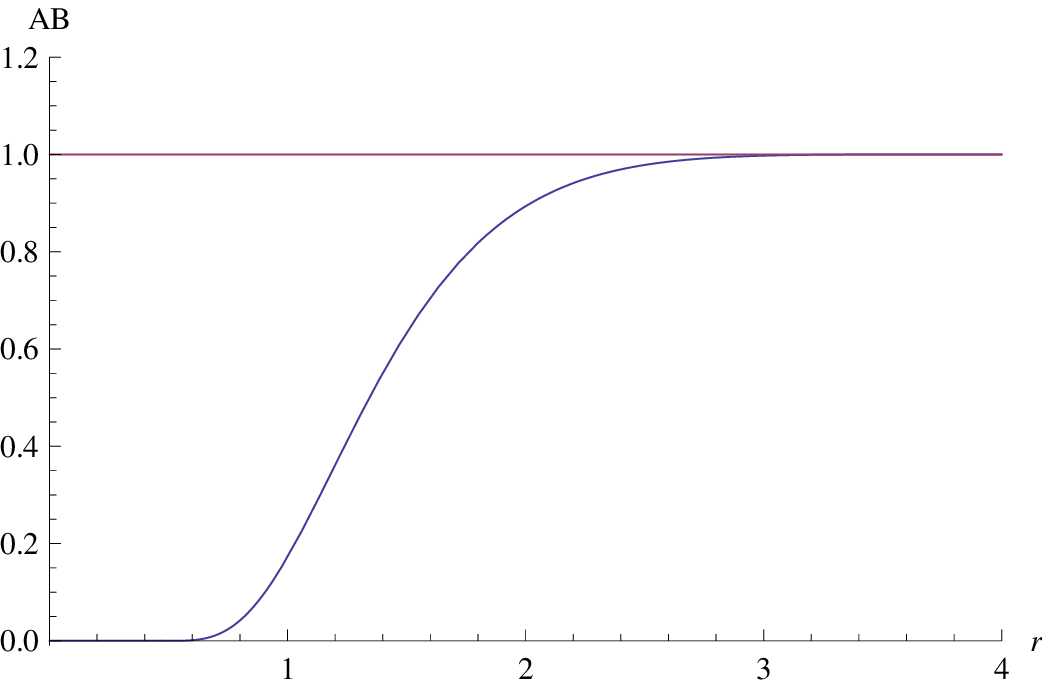}
\caption{Singular shell case. Product $A(r)B(r) $ for $\lambda =1$ and $\kappa = 0.1$. Shooting from the compacton boundary for the initial values $m_s = 0.5$ and $R=4$. The resulting singular boundary is at $r_0 = 0.4736$ (position of the zero of $AB$). }
\end{center} 
\end{figure}

We show two cases in Figures 18 - 25, where we choose $R_s \equiv 2m_s =1$ in both cases.
The corresponding small compacton radius is $R_{sc} = 4.885$, and we choose $R=4$ in Figs. 17 - 21, and $R=2$ in Figs. 22 - 25. For a value of $R$ near $R_{sc}$ (the first case $R=4$), the metric functions deviate form the Schwarzschild behaviour rather soon, and the transition from $r=R$ to $r=r_0$ is rather mild. For a $R$ near $R_s$, on the other hand, the metric functions $B$ and $A$ behave almost exactly like the Schwarzschild solution for most values of $r$. This is clearly seen in the plot of the product $AB$ in Figure 25. Specifically, $B$ almost reaches the Schwarzschild horizon divergence (in the concrete example for $R=2$, it has a maximum value of $B_{\rm max} >500$), before steeply descending to zero, see Fig. 23. 

We shall find in the next section that the condition that $B(r_0)=0$ at the inner shell boundary $r=r_0$ indeed implies that some curvature invariants become singular there, justifying the name "singular shell". In fact, space-time itself ends at the singular boundary, i.e., cannot be continued beyond the singularity (remember that we found in Section 3.1 that there is no continuation of $B(r)$ to values $r<r_0$ if $B(r_0)=0$). We will also find, in the next section, that the singularity is, at the same time, the locus of a Killing horizon, and no signals can propagate from the singular surface to an outside observer, i.e., the singularity is invisible. In this sense, these singular shells have some similarity with ordinary black holes.    

\begin{figure}[h!]
\begin{center}
\includegraphics[width=0.65\textwidth]{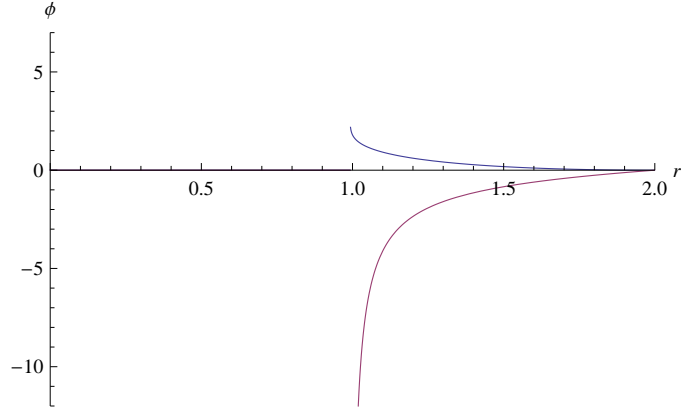}
\caption{Singular shell case. Functions $\phi(r)$ and $\phi '(r)$ for $\lambda =1$ and $\kappa = 0.1$. Shooting from the compacton boundary for the initial values $m_s = 0.5 $ and $R=2$. The resulting singular boundary is at $r_0 = 0.9928$ (position of the minimum of $\phi '$, which is outside the range of the plot). }
\end{center} 
\end{figure}

\begin{figure}[h!]
\begin{center}
\includegraphics[width=0.65\textwidth]{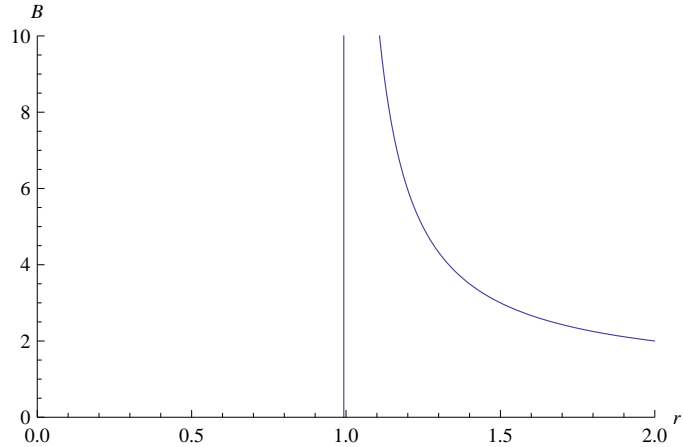}
\caption{Singular shell case. Function $B(r)$ for $\lambda =1$ and $\kappa = 0.1$. Shooting from the compacton boundary for the initial values $m_s = 0.5$ and $R=2$. The resulting singular boundary is at $r_0 = 0.9928$ (position of the zero of $B$). The maximum of $B$ is outside the range of the plot, $B_{\rm max} > 500$.  }
\end{center} 
\end{figure}

\begin{figure}[h!]
\begin{center}
\includegraphics[width=0.65\textwidth]{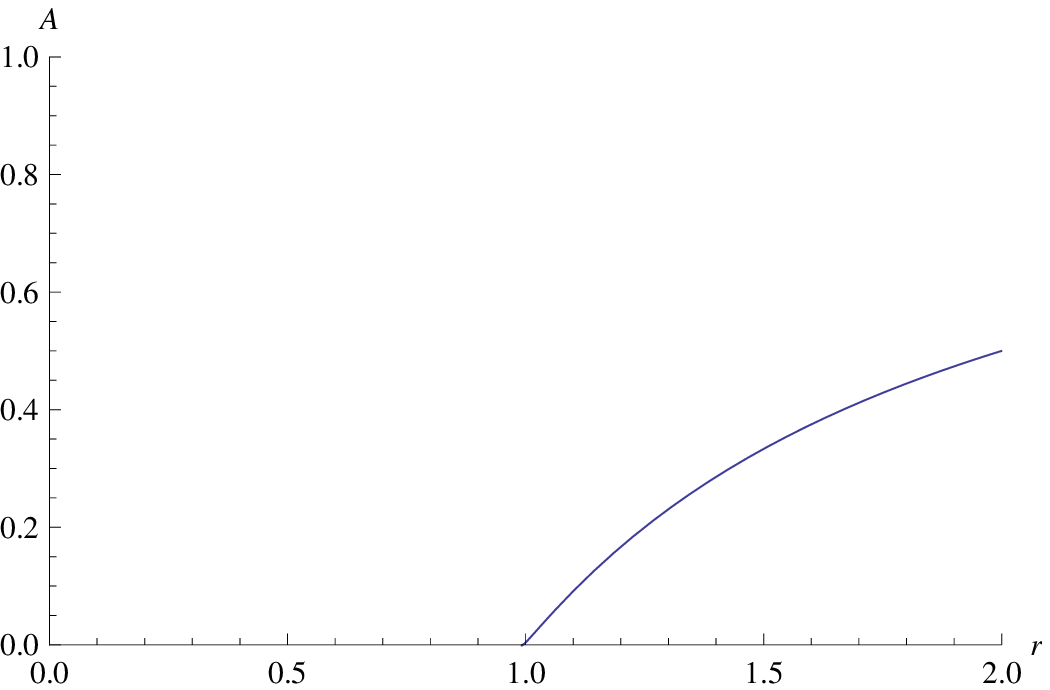}
\caption{Singular shell case. Function $A(r)$ for $\lambda =1$ and $\kappa = 0.1$. Shooting from the compacton boundary for the initial values $m_s = 0.5$ and $R=2$. The resulting singular boundary is at $r_0 = 0.9928$ (position of the zero of $A$).   }
\end{center} 
\end{figure}

\begin{figure}[h!]
\begin{center}
\includegraphics[width=0.65\textwidth]{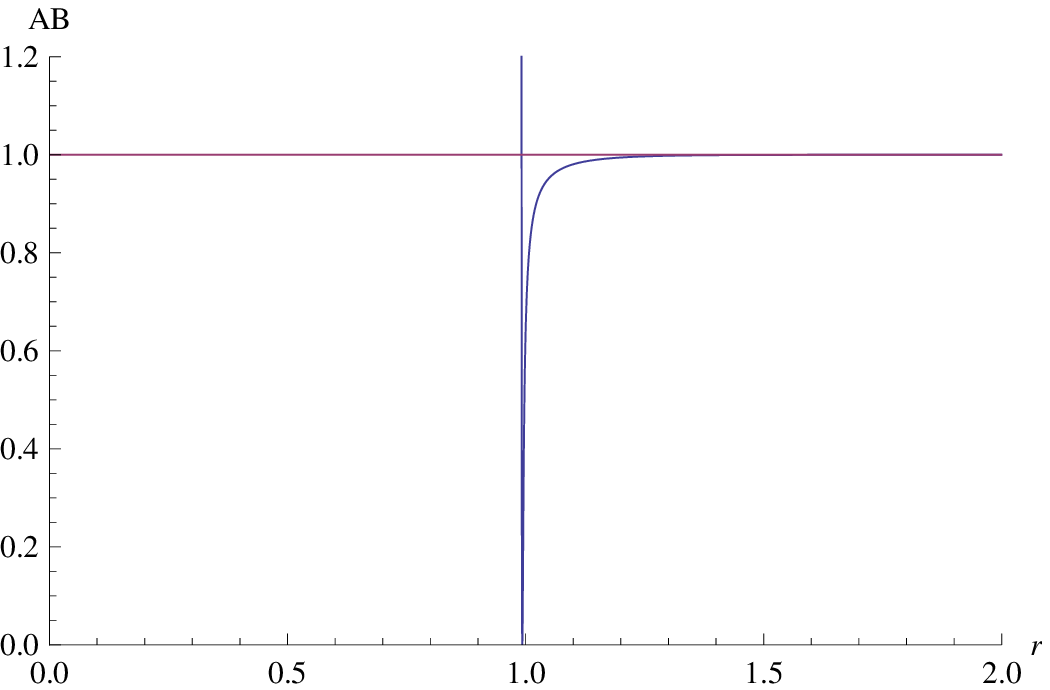}
\caption{Singular shell case. Product $A(r)B(r) $ for $\lambda =1$ and $\kappa = 0.1$. Shooting from the compacton boundary for the initial values $m_s = 0.5$ and $R=2$. The resulting singular boundary is at $r_0 = 0.9928$ (position of the zero of $AB$).  }
\end{center} 
\end{figure}

\subsection{Singularities of curvature invariants and radial geodesics}  

We found two types of solutions in the last sections, namely compact balls and singular shells.
Within the compact balls, solutions are further discriminated by the approach of $B$ and $A$ to the center $r=0$. The first case of large compactons is the case $\bar b_1 >0$, where $B$ approaches the center linearly, $B \sim \bar b_1 r$, and $A\sim \bar a_{-1} r^{-1}$.
The second case of small compactons is defined by  $\bar b_1 =0$ and, as a consequence, $B \sim \bar b_2 r^2$ and $A \sim \bar a_2 r^2$ (here $\bar b_2$ is, in fact, not a free parameter, but this issue is irrelevant for the present discussion). 

Both types of solutions have the property that the Riemann tensor and Ricci tensor are singular at $r=0$. 
Singularities show their presence in a most physically relevant way in invariants which may be derived from the Riemann and Ricci tensors. A further simplification occurs for invariants derivable form the Ricci tensor alone, because then the Einstein equations may be used.
Concretely we find for the Ricci scalar
\begin{equation}
{\cal R} = -\kappa^2 T^\mu_\mu = 4 \kappa^2 \lambda \phi^2
\end{equation}
so the Ricci scalar is, in fact, regular at the origin or the singular boundary, because the scalar field $\phi$ is. The Ricci tensor squared may be expressed like
\begin{equation}
{\cal R}_{\mu\nu}{\cal R}^{\mu\nu} = \kappa^4 \left( T_{\mu\nu}T^{\mu\nu} + (T_\mu^\mu)^2
\right) = \kappa^4 \left( \frac{3}{4}\frac{\phi'^8}{B^4} + 20 \lambda^2 \phi^4 \right)
\end{equation}
where we used
\begin{equation}
T_{\mu\nu}T^{\mu\nu}  = \frac{3}{4}\frac{\phi'^8}{B^4} + 4\lambda^2 \phi^4 .
 \end{equation}
Here, the first term is singular at $r=0$, because $\phi'$ is regular and non-zero there whereas $B$ goes to zero like $r$ (large compacton),  $r^2$ (small compacton), or $(r-r_0)$ (singular shell), respectively. Therefore, the Ricci tensor squared is already singular. More complicated invariants like, e.g., the Kretschmann invariant, are singular, as well. 
 
One possible question to be asked in connection with these singularities is whether they really belong to the space time manifold, that is, whether a freely falling particle may reach them in finite proper time.  We will find in all three cases that this is the case, i.e. there exist geodesics which hit the singularities in finite proper time. For this purpose we have to study the geodesic equation
\begin{equation}
\ddot x^\mu (\tau) + \Gamma^\mu_{\alpha \beta} \dot x^\alpha \dot x^\beta =0.
\end{equation}
Here $\Gamma^\mu_{\alpha \beta}$ is the Christoffel connection and $\tau$ is the proper time. We shall restrict to radial geodesics, which is sufficient for our purpose. In this case the geodesic equations reduce to
\begin{eqnarray}
\ddot t + \Gamma^t_{tt} \dot t^2 + 2 \Gamma^t_{tr} \dot t \dot r + \Gamma^t_{rr} \dot r^2
&=& 0 \nonumber \\
\ddot r + \Gamma^r_{tt} \dot t^2 + 2 \Gamma^r_{tr} \dot t \dot r + \Gamma^r_{rr} \dot r^2 &=&0
\end{eqnarray}
where the nonzero components of the Christoffel connection are
\begin{eqnarray}
\Gamma^t_{tr} &=& \frac{1}{2} g^{tt}\partial_r g_{tt} = \frac{1}{2} \frac{A'}{A} \nonumber \\
\Gamma^r_{tt} &=& -\frac{1}{2} g^{rr}\partial_r g_{tt} = \frac{1}{2} \frac{A'}{B} \nonumber \\
\Gamma^r_{rr} &=& \frac{1}{2} g^{rr}\partial_r g_{rr} = \frac{1}{2}\frac{B'}{B}
\end{eqnarray}
We are interested in the geodesic motion near the singularity, therefore it is enough for our purposes to restrict to the leading behaviour of $A$ and $B$. In the large compacton case , this leading behaviour is $A \sim (\bar a_{-1}/r)$ and $B\sim \bar b_1 r$ and we get the geodesic equations
\begin{eqnarray}
\ddot t - \frac{1}{r} \dot t \dot r &=& 0 \nonumber \\
\ddot r - \frac{\gamma}{2r^3} \dot t^2 + \frac{1}{2r} \dot r^2 &=&0
\end{eqnarray}
where $\gamma \equiv (\bar a_{-1}/\bar b_1 )$. The equation for $t$ has the solution
\begin{equation}
t = c_1 \int_{\tau_0}^\tau d\tau ' r(\tau ') + c_2
\end{equation}
and the equation for $r$ becomes
\begin{equation}
\ddot r - \frac{c_3}{r} + \frac{1}{2r}\dot r^2 =0
\end{equation}
where $c_3 \equiv (\gamma c_1^2 /2)>0$. Under the transformation $r = f^\frac{2}{3}$ the last equation turns into
\begin{equation}
\ddot f = \frac{3c_3}{2} f^{-\frac{1}{3}} .
\end{equation}
This last equation has an easy interpretation in terms of an equivalent mechanical system. It is just the e.o.m. of a nonrelativistic particle in one dimension in the repulsive external potential $U(f)\sim -f^\frac{2}{3}$. The center $f=0$ of the potential may be reached despite its repulsive nature because of the positive power $(2/3)$ in the distance law. However, not all radial geodesics hit the center $r=0$, because in the equivalent mechanical problem the particle needs a sufficient initial velocity.

The small compacton case may be analysed in an equivalent manner. The leading
behaviour is $A \sim \bar a_2 r^2$ and $B\sim \bar b_2 r^2$ and we get the geodesic equations
\begin{eqnarray}
\ddot t + \frac{2}{r} \dot t \dot r &=& 0 \nonumber \\
\ddot r + \frac{\gamma}{r} \dot t^2 + \frac{1}{r} \dot r^2 &=&0
\end{eqnarray}
where $\gamma \equiv (\bar a_2/\bar b_2 )$. The equation for $t$ has the solution
\begin{equation}
t = c_1 \int_{\tau_0}^\tau \frac{d\tau '}{ r^2(\tau ')} + c_2
\end{equation}
and the equation for $r$ becomes
\begin{equation}
\ddot r + \frac{c_3}{r^5} + \frac{1}{r}\dot r^2 =0
\end{equation}
where $c_3 \equiv (\gamma c_1^2 /2)>0$. Under the transformation $r = f^\frac{1}{2}$ the last equation turns into
\begin{equation}
\ddot f = -\frac{2c_3}{f^2}  .
\end{equation}
This is the e.o.m. of a nonrelativistic one-dimensional particle in an attractive $1/f$ potential.
The particle will hit the center $f=0$ unless it has a sufficiently large (outward directed) initial escape velocity. 

Finally, for the singular shell it is useful to introduce the variable $u=r-r_0$. Then the leading
behaviour is $A \sim \bar a_3 u^3$ and $B\sim \bar b_1 u$ and we get the geodesic equations
\begin{eqnarray}
\ddot t + \frac{3}{u} \dot t \dot u &=& 0 \nonumber \\
\ddot u+ \frac{3\gamma u}{2} \dot t^2 + \frac{1}{2u} \dot u^2 &=&0
\end{eqnarray}
where $\gamma \equiv (\bar a_3/\bar b_1 )$. The equation for $t$ has the solution
\begin{equation}
t = c_1 \int_{\tau_0}^\tau \frac{d\tau '}{ u^3(\tau ')} + c_2
\end{equation}
and the equation for $u$ becomes
\begin{equation}
\ddot u + \frac{c_3}{u^5} + \frac{1}{2u}\dot u^2 =0
\end{equation}
where $c_3 \equiv (3\gamma c_1^2 /2)>0$. Under the transformation $u = f^\frac{2}{3}$ the last equation turns into
\begin{equation}
\ddot f = -\frac{3c_3}{2f^3}  .
\end{equation}
This is the e.o.m. of a nonrelativistic one-dimensional particle in an attractive $1/f^2$ potential.
The particle will hit the center $f=0$ unless it has a sufficiently large (outward directed) initial escape velocity.

We conclude that in all cases there exist radial geodesics which hit the singularities at $r=0$ or $r=r_0$, respectively, therefore these singularities belong to the corresponding space time manifolds.

The singular shells, however, differ from the compact balls in one essential aspect, in that their singular inner boundary is, at the same time, the locus of a Killing horizon and, therefore, invisible for an outside observer. In a first step, let us consider the equation for a light-like radial geodesic. We find
\begin{equation}
ds^2 = 0 = -Adt^2 +Bdr^2 \quad \Rightarrow \quad dt = \sqrt{\frac{B}{A}}dr
\end{equation}
and with $B/A \sim (c/(r-r_0))^2$ near $r=r_0$ we find the solution
\begin{equation}
r(t) - r_0 = \exp \left( \frac{t-t_0}{c} \right)
\end{equation}
where $t_0$ is an integration constant. Obviously, $r=r_0$ requires $t=-\infty$, so a light ray cannot escape from the singular surface. A slightly more rigorous and less coordinate dependent derivation uses the concept of a Killing horizon. Let us assume the existence of a Killing vector $\xi$ and consider the set of points (hyper-surface) $H_0$ where $\xi$ is null, i.e., $N\equiv (\xi ,\xi) =0$. A Killing horizon $H$ is a connected component $H \in H_0$ which is, at the same time, a null hyper-surface, i.e., $M\equiv (dN,dN)=0$. So let us demonstrate that the inner boundary of the singular shell is indeed a Killing horizon. For the Killing vector we choose the generator of time translations $\xi = (\partial / \partial t) \equiv \partial_t$, as is obvious for a static space-time. For $N$ we find
\begin{equation}
N\equiv (\xi ,\xi ) =g_{\mu\nu}dx^\mu \otimes dx^\nu ( \partial_t ,\partial_t ) =-A .
\end{equation}
As near $r=r_0$ $A$ behaves like $A \sim \bar a_3 (r-r_0)^3$, it holds that $N(r_0)=0$. For $M$ we get with $dN = -A'dr$
\begin{equation}
M\equiv (dN,dN) = A'^2 g^{\mu\nu}\partial_\mu \otimes \partial_\nu (dr,dr) = \frac{A'^2}{B} .
\end{equation}
With the leading behaviour of $A$ near $r_0$ like above and $B\sim \bar b_1 (r-r_0)$, we easily conclude that indeed $M(r_0)=0$, so the singular boundary is a Killing horizon. We remind the reader that for static space-times every event horizon is a Killing horizon, which makes our result all the more interesting.  The singular shell solutions we found are, in fact, quite similar to ordinary black holes in this respect.

\section{Discussion}
It has been the main purpose of the present article to investigate in detail the properties of compact boson stars (compact balls or shells
minimally coupled to gravity) for a theory with a non-standard kinetic term in three plus one dimensions. 
We chose the simplest possible theory for this purpose, with a purely quartic kinetic term and the simple quadratic potential, because we wanted to pursue the analytical investigation as far as possible in order to understand also more generic and qualitative features, in addition to  explicit numerical calculations.  This theory has a one-parameter family of compact ball solutions already in the case without gravity. The compact balls in the theory without gravity turn out to be solutions of the weak type, which solve the field equations everywhere except at the origin. We remark that this fact is not a problem in the case at hand, because the set of weak solutions is the appropriate solution set for variational problems of this type, in any case. 
Alternatively, the compact ball solutions may be extended to solutions in the whole space by introducing a delta function source term at the origin. It turned out that these solutions may have arbitrary size and energy.  

In the case with gravity, we found two different types of solutions, both of which are of the strong type, i.e., they solve the field equations everywhere. For compact balls, we found that the scalar field itself is behaving well at the origin (it has a power series expansion there). The metric coefficients do have a power series expansion at the origin, as well, (with the exception of a $r^{-1}$ leading term in one specific case), but nevertheless give rise to a singularity at $r=0$. Specifically, some higher invariants of the Riemann tensor, like the Ricci tensor squared, or the Kretschmann invariant, become singular at this point.  This singularity is not shielded by any horizon, so it is a naked singularity. In spite of the naked singularity at the center,  asymptotically these solutions behave like Schwarzschild solutions, so they have finite Schwarzschild mass (or asymptotic, ADM mass) $m_s$. In one class of solutions (the large compactons), this ADM mass may take on negative values, in spite of the positive definite energy density. The theorems excluding this type of behaviour are evaded by the presence of naked singularities in the case at hand.   

In addition, we found another type of solutions in the theory with gravity, namely the singular shells where a singularity already forms at a finite, nonzero value of the radial coordinate. 
This singularity is a surface in space with finite area, as may be checked easily by inspecting the corresponding metric (the angular part of the metric is not multiplied by any suppressing factor at the position $r=r_0$ of the singularity). Further, 
this singularity is a genuine singular boundary of space-time,
because space-time cannot be continued beyond this boundary. At the same time, this singular boundary is a Killing horizon.  
In addition, the singular shells asymptotically for large radius are Schwarzschild, 
 so they resemble ordinary Schwarzschild black holes in many respects. Certainly they are black (i.e., no radiation may escape from the horizon), although the curvature near the horizon is different from the Schwarzschild black hole case. It is probably more a question of parlance whether these objects should be called genuine black holes - thereby providing a counter-example to the scalar no hair conjecture - or interpreted as different object which, nevertheless, share many properties with ordinary black holes. In any case, the existence of these singular shell solutions is interesting. On the other hand, 
ordinary Schwarzschild black hole solutions (where the curvature near the horizon behaves like in the Schwarzschild case) do not exist in our model, except for the trivial case where the metric is exactly Schwarzschild everywhere and the scalar field takes its vacuum value in all space-time. 

At this point, an obvious question to ask is whether theories of the type presented here may be of some relevance in astrophysical or cosmological contexts. The theory presented here is the simplest possible case of the class of theories having a non-standard kinetic term and allowing for compact boson stars. Therefore, an attempt for direct applications would most likely be premature, and we shall focus, instead, on more generic issues.
Specifically, we want to comment on some features of these theories which might make them interesting for astrophysical considerations. A first feature we already mentioned is the fact that non-overlapping multi compact ball configurations do not interact at all in the case without gravity, so the self-gravitating ones only interact via the universal gravitational interaction. 
A further feature of compact solitons in K field theories in general is the fact that linear fluctuations are possible only inside the compact solitons, whereas they are completely suppressed in the vacuum. This fact is easy to understand. Unlike for the standard kinetic term, in the case of a non-standard K theory the wave operator acting on the fluctuation field $\delta \phi$ is multiplied by some power of $X\equiv \partial^\mu \phi \partial_\mu \phi$, where $X$ has to be evaluated at the background field (i.e., the compacton). But $X$ is identically zero outside the compacton (that is, in the region where the background field takes its vacuum value), which is the reason for the complete suppression of linear fluctuations in that region.    
This implies that a compacton in K field theories cannot emit particle type radiation, which makes it a rather stable object. These two facts - that compact boson stars only interact gravitationally, and that particle-like excitations in vacuum do not exist - might make compact boson stars of the type discussed in this paper interesting candidates for dark matter. The second fact may, further, give rise to the following speculations. Let us assume for a moment that a universe with a scalar K field theory was a reasonably good effective description in a certain (early) stage of the evolution of the universe. Then linear fluctuations of the K field which may act as seeds for matter formation were naturally present in the compacton regions, but absent in the vacuum regions. The late time remnants of the former would be regions of the universe with a higher matter concentration, whereas the latter might have evolved into voids (low matter density regions). 

Before more convincing conclusions can be drawn with respect to possible applications of compact boson stars, one key issue to understand is which features of the theory presented in this article are generic and which are just artifacts of the specific model. One important feature in this respect obviously is the presence of naked singularities in the compact ball type  solutions of the model studied in this paper. Such naked singularities of the point-like type are already present in finite mass solutions (boson stars) for standard self-gravitating scalar fields \cite{Buchdahl:1959nk} - \cite{Xanthopoulos:1989kb}. Further, it is known that the gravitational collapse of a standard scalar field may terminate in a naked singularity \cite{Christodoulou:1994hg}, \cite{Dwivedi:1994qs}. Nevertheless, the presence of naked singularities is frequently viewed as an unwanted or even unphysical feature and, in any case, the generic or non-generic nature of naked singularities in compact balls is an interesting question. We cannot give a completely mathematically rigorous answer to this question, but shall try to partially answer it relying on our experience and some recent results in the literature.  On the one hand, for a static, radially symmetric scalar field, it seems to be a typical feature that non-singular metric functions $A$, $B$ at the origin $r=0$ require that the scalar field takes its vacuum value there. Further, these local solutions typically are isolated solutions (i.e., without free integration constants) which makes it difficult to connect them to finite mass (i.e., Schwarzschild) asymptotic solutions. In our specific model, we could exclude this possibility, whereas in more general theories it might still be possible to connect a regular center with an asymptotic Schwarzschild solution provided that {\em i)} the scalar field has more than one vacuum and {\em ii)} the scalar field Lagrangian contains sufficiently many coupling constants which may be varied in order to connect the two local solutions near zero and near infinity.  

There exists, however, a class of slightly different but related theories where the appearance of compact balls with a regular center seems to be a more generic feature, namely the class of self-gravitating compact Q balls. A Q ball is a complex scalar field $\psi$ restricted to the ansatz  $\psi = \exp (i\omega t) \phi (\vec x)$, such that the resulting static equations for the real function $\phi$ are quite similar to the static field equations of a real scalar field, both with and without gravitation. 
Q balls (without gravitation) were introduced by S. Coleman in \cite{Coleman:1985ki} and have been widely investigated since then. Systems of self-gravitating Q balls are frequently subsumed under the general notion of boson stars. For more details we refer to the review articles  \cite{Jetzer:1991jr} - \cite{Schunck:2003kk}.
Compact Q balls (without gravity) for a complex scalar theory with a standard kinetic term but with a V-shaped potential have recently been found in \cite{Arodz:2008jk} and in \cite{Arodz:2008nm} (the second paper included also an electromagnetic interaction of the complex scalar, and allows both for ball-shaped and shell-shaped solutions). The gravitational counterpart of the model studied in \cite{Arodz:2008nm} was investigated in 
\cite{Kleihaus:2009kr}, mainly using numerical methods, and the existence of non-singular solutions (both of the ball type and of the shell type) was established. Therefore, the existence of regular self-gravitating compact solutions seems to be a more generic feature for Q balls, analogously to the case of standard (non-compact) Q balls.    

In any case, we think that the further study of the interaction between compact objects (like compact solitons, Q balls, etc.) and gravitation is an interesting endeavour which may lead to new insights and unexpected phenomena, as well as some applications to astrophysics and cosmology. One future line of investigation will certainly be the study of Q balls. Another promising direction is the analysis of the symmetries and conservation laws of the corresponding theories, e.g., along the lines of \cite{Alvarez:1997ma} - \cite{Adam:2006vb}, which may lead to a better understanding of more general solutions as well as their stability. These issues are under current investigation.     

\section*{Acknowledgements}
C.A., P.K. and J.S.-G. thank MCyT (Spain) and FEDER
(FPA2005-01963), and support from
 Xunta de Galicia (grant PGIDIT06PXIB296182PR and Conselleria de
Educacion). A.W. acknowledges support from the Foundation for
Polish Science FNP (KOLUMB programme 2008/2009) and Ministry of
Science and Higher Education of Poland grant N N202 126735
(2008-2010). N.E.G. is partially supported by the ANPCyT Grant PICT-2007-00849. He was visiting ICTP under the ICTP associate program during the late stages of this work, and wants to thank the ICTP for providing support and an extremely comfortable work environment.

\end{document}